\documentclass[aps,prd,eqsecnum,showpacs,superscriptaddress,preprint,
floatfix,preprintnumbers,altaffilletter]{revtex4}
\usepackage{graphicx}
\usepackage[colorlinks=true]{hyperref}
\usepackage{amssymb}
\usepackage{amsmath}
\usepackage{longtable}
\usepackage{rotating}
\usepackage{color}
\usepackage{epsfig}
\usepackage{units}
\begin{document}

\title{
  Probing the anisotropies of a stochastic gravitational-wave background using a network of ground-based laser interferometers
}

\author{Eric~Thrane}
\email{ethrane@physics.umn.edu}
\affiliation{School of Physics and Astronomy,
University of Minnesota, Minneapolis, MN 55455, USA}

\author{Stefan~Ballmer}
\email{sballmer@caltech.edu}
\affiliation{LIGO Laboratory, California Institue of Technology, MS 18-34, 
Pasadena, CA 91125, USA}

\author{Joseph~D.~Romano}
\email{joe@phys.utb.edu}
\affiliation{Department of Physics and Astronomy, 
The University of Texas, Brownsville, Texas 78520, USA}

\author{Sanjit~Mitra}
\email{smitra@ligo.caltech.edu}
\affiliation{Jet Propulsion Laboratory, California Institute of Technology, Pasadena, CA 91109, USA}
\affiliation{LIGO Laboratory, California Institue of Technology, MS 18-34, 
Pasadena, CA 91125, USA}
\affiliation{Observatoire de la C\^ote dÕAzur, BP 4229, 06304 Nice Cedex 4, France}

\author{Dipongkar~Talukder}
\email{talukder\_d@wsu.edu}
\affiliation{Department of Physics,
Washington State University, Pullman, WA 99164-2814, USA}

\author{Sukanta~Bose}
\email{sukanta@mail.wsu.edu}
\affiliation{Department of Physics,
Washington State University, Pullman, WA 99164-2814, USA}

\author{Vuk~Mandic}
\email{mandic@physics.umn.edu}
\affiliation{School of Physics and Astronomy,
University of Minnesota, Minneapolis, MN 55455, USA}

\date{\today}

\begin{abstract}
We present a maximum-likelihood analysis for estimating the angular
distribution of power in an anisotropic stochastic
gravitational-wave background using ground-based laser interferometers.
The standard isotropic
and gravitational-wave radiometer searches
(optimal for point sources)
are recovered as special limiting cases.
The angular distribution can be decomposed with respect to {\em any}
set of basis functions on the sky, and the single-baseline,
cross-correlation analysis is easily extended to
a network of three or more detectors---that is, to multiple baselines.
A spherical harmonic decomposition, which provides
maximum-likelihood estimates of the multipole moments of the
gravitational-wave sky, is described in detail.
We also discuss:
(i) the covariance matrix of the estimators and its relationship
to the detector response of a network of interferometers,
(ii) a singular-value decomposition method for regularizing the
deconvolution of the detector response from the measured
sky map,
(iii) the expected increase in sensitivity obtained by including multiple
baselines, and
(iv) the numerical results of this method when applied to simulated
data consisting of both point-like and diffuse sources.
Comparisions between this general method and the standard isotropic
and radiometer searches
are given throughout, to make contact with the existing literature on
stochastic background searches.
\end{abstract}

\preprint{LIGO-P0900083}

\maketitle

\section{INTRODUCTION}
\label{s:intro}

Data from the laser interferometric gravitational-wave detectors
LIGO \cite{LIGO, Barish:1999,LIGO_eprint,detector},
Virgo \cite{Virgo, Bradaschia:1990}, 
and GEO \cite{GEO-600, Wilke:2004}
are currently being analysed for the presence of 
gravitational waves
from a variety of sources.
These include 
signals from inspiraling and coalescing 
compact binaries 
(for example, neutron stars and/or stellar mass 
black holes)~\cite{bin-insp-S3S4,lowmass_S5,ringdowns_S4,lowmass_S5B},
continuous gravitational waves from quasi-periodic
sources such as pulsars~\cite{SGR,SGR1806,radiopulsars,allsky},
and bursts of gravitational radiation 
associated with gamma-ray bursts~\cite{GRB070201,39GRBs,S4GRB}, core-collapse
supernovae, or other violent events~\cite{S3bursts}.
In addition, searches are ongoing for the presence
of a background of {\em stochastic} gravitational 
radiation of either astrophysical or cosmological 
origin~\cite{radiometer,LIGO-ALLEGRO,S4Isotropic}, whose 
detection might provide insights about 
the very early universe \cite{Maggiore:2000}, 
well before the production of 
the cosmic microwave background
\cite{Kolb:1999}.

Although no direct detections of gravitational waves have been made to date, the most recent data taken are of unprecented sensitivity~\cite{stoch-S5,SGR,GRB070201}, leading to upper limits on gravitational-wave strengths that are competitive with or surpass those from electromagnetic or particle physics observations.
Of particular note is the upper limit on 
the strength of a gravitational-wave signal 
from the 
Crab pulsar \cite{LIGO-crab}, which is a
factor of 1.6 lower than the corresponding 
limit inferred from electromagnetic 
pulsar spin-down observations~\cite{palomba}.
Also, the current direct limit on the strength of an isotropic stochastic gravitational-wave background at $\unit[100]{Hz}$ $\Omega_{\text gw}<6.9\times10^{-6}$~\cite{stoch-S5} (at 95\% confidence) has surpassed bounds set by considerations from Big Bang Nucleosynthesis~\cite{Maggiore:2000} and from the microwave background~\cite{CMB-limit}.
 
In this paper, we describe an analysis 
method that estimates the 
angular distribution of power in an
{\em anisotropic}
stochastic gravitational-wave background.  
This method includes both the standard isotropic
\cite{S1Isotropic,S3Isotropic,S4Isotropic}
and gravitational-wave radiometer 
searches \cite{mitra-et-al,radiometer}
(optimal for point sources) 
as special limiting cases.
(For our purposes {\em anisotropic} is taken 
to mean {\em not necessarily isotropic}.)
Similar to the radiometer technique, 
the method 
presented here looks for modulations in the 
gravitational-wave signal induced by the
Earth's rotational motion relative to an
anisotropic background.
The method provides maximum-likelihood 
estimates of the angular distribution of 
gravitational-wave power 
${\cal P}(\hat\Omega)=\sum_\alpha {\cal P}_\alpha{\bf e}_\alpha(\hat\Omega)$, 
decomposed with respect 
to some set of basis functions on the sky.
By choosing a pixel basis
${\bf e}_{\hat\Omega'}(\hat\Omega)=\delta(\hat\Omega,\hat\Omega')$,
we recover the results of the radiometer 
method discussed in~\cite{mitra-et-al,radiometer}.
By choosing the spherical harmonics basis
$Y_{lm}(\hat\Omega)$
defined with respect to the Earth's 
rotational axis, 
we obtain maximum-likelihood estimates of the 
{\em multipole moments} 
${\cal P}_{lm}$ of the 
gravitational-wave sky.
This basis is particularly convenient as
the standard isotropic analysis 
corresponds to simply restricting attention 
to the monopole moment ${\cal P}_{00}$, while the point-source 
radiometer results are well-approximated by
choosing a sufficiently large value of $l_{\rm max}$ 
($l_{\rm max}\sim30$), appropriate for the 
diffraction-limited beam pattern at $f\sim 1$~kHz.
In addition, the use of spherical harmonics 
simplifies the problem of removing the 
`smearing' effects of the beam pattern
from the measured sky map 
(that is, deconvolution of the {\em dirty map}), 
given the smaller number of elements and symmetries 
of the beam pattern matrix with respect to 
the $lm$ indices.
The problem of deconvolving a cross-correlated gravitational-wave signal from the interferometers' beam pattern in the spherical harmonic basis has been described in~\cite{Cornish:2001hg}.
We address this problem in detail in section~\ref{s:data_analysis}.
We further note that the spherical harmonic basis is useful for
the efficient analysis of cross-correlated data in a variety of applications
including searches for transient gravitational-wave sources~\cite{kipp}.

Regardless of basis, the method described here 
is easily extended to work with a network 
or three or more detectors with uncorrelated 
detector noise, by simply adding the individual 
baseline beam patterns and dirty maps before
deconvolution.
A multi-baseline analysis improves the overall
sensitivity of the search by reducing the 
variances of the individual estimators, and 
provides a natural way of regularising the 
deconvolution of the dirty map;
the beam pattern matrix has fewer
null (or nearly null) directions for multiple
baselines and is thus more stable during inversion.

The structure of the rest of the paper is the following:
In section~\ref{s:sgwb}, we briefly review the
statistical properties of an anisotropic background,
and show how a generalized overlap reduction 
function arises in a cross-correlation search for
such a background.
In section~\ref{s:MLestimation} we derive the optimal estimators 
of the angular distribution of the gravitational-wave 
power, starting from the likelihood
function for cross-correlated data.
We explicitly construct the beam pattern matrix, 
and discuss its relation to the covariance matrix
of the estimated ${\cal P}_{\alpha}$.
Section \ref{s:data_analysis} describes details of
the data analysis 
implementation and issues related to deconvolution
and regularisation.
It also
briefly describes how
to extend 
the analysis
to a network of three or more detectors, and the expected 
increase in sensitivity from using multiple baselines.
In section~\ref{s:simulations} we present numerical 
results of 
the method applied to simulated data.
We consider both point-like and diffuse-source 
injections, and compare the extracted and injected 
sky-maps.
Finally, in section~\ref{s:summary} we summarize our results.
We also include three appendices:
Appendices ~\ref{s:Ylms} and \ref{s:identities}
contain
definitions of the spherical harmonics and 
some useful identities relating different 
multipole moments,
beam pattern matrix components, etc.;
Appendix \ref{s:DetStat} defines a related 
detection statistic that assumes a particular
distribution of (normalized) angular distribution 
functions on the sky.

\section{GRAVITATIONAL-WAVE BACKGROUNDS}
\label{s:sgwb}

Stochastic gravitational-waves are produced by the 
superposition of a large number of weak, 
independent, unresolved gravitational-wave sources.
The signal can be either cosmological or 
astrophysical in
nature, leading to different expected characteristics:
(i) A cosmological background, consisting e.g.,  
of remnant gravitational waves left over from the very 
early universe, is expected to be predominantly 
isotropic, similar to that of the 2.73~K temperature 
distribution of the cosmic microwave background. 
(ii) An astrophysical background, on the other hand, 
produced by more recent astrophysical events,
such as early-phase compact binary inspiral or 
continuous radiation from pulsars (see, e.g., 
\cite{VukTania}), 
will most likely be {\em anisotropic}, 
following the spatial distribution of the sources.
In addition, cosmological backgrounds are expected 
to have relatively smooth, monotonic power spectra
(for example, falling off close to $f^{-3}$ for standard
inflationary models)~\cite{Boyle:2005se}, while astrophysical backgrounds 
are expected to have power peaked at some characteristic
frequency.
Not surprisingly,
to optimally search for these different signals 
requires different search algorithms, adapted for 
the angular distribution and 
spectral properties of the source.
In this section, we describe how the anistropy 
of a stochastic gravitational-wave background 
manifests itself in the 
statistical properties of the signal, and in the 
expected value of the 
cross-correlation of the output of two detectors.
The following sections 
then describe how one can search for such a signature 
in the measured data.

\subsection{Statistical properties}
\label{s:statproperties}

In the transverse-traceless gauge,
the metric perturbations due to a stochastic gravitational-wave
background can be written as a superposition of plane waves
having frequency $f$ and propagating in the direction $\hat\Omega$:
\begin{equation}
h_{ab}(t,\vec x)
=
\int_{-\infty}^\infty df
\int_{S^2} d\hat\Omega\,
e^A_{ab}(\hat\Omega)
h_A(f,\hat\Omega)\,
e^{i 2\pi f(t-\hat{\Omega}\cdot \vec x/c)}
\label{e:hab}
\end{equation}
where $e^A_{ab}(\hat\Omega)$ are the 
gravitational-wave 
polarization tensors.
(Summation over polarization indices $A$ is
understood.)
In standard angular coordinates on the two-sphere
$\theta\in[0,\pi]$, $\phi\in[0,2\pi)$,
we can write
\begin{eqnarray}
\hat\Omega
&=&
\sin\theta\cos\phi\,\hat x+
\sin\theta\sin\phi\,\hat y+
\cos\theta\,\hat z
\,,
\\
\hat l
&=&
\cos\theta\cos\phi\,\hat x+
\cos\theta\sin\phi\,\hat y-
\sin\theta\,\hat z
\,,
\\
\hat m
&=&
-\sin\phi\,\hat x+
\cos\phi\,\hat y
\,,
\end{eqnarray}
so that $\{\hat l,\hat m,\hat \Omega\}$ forms a
right-handed system of unit vectors.
We can then define the two ($A=+,\times$) polarization 
tensors to be
\begin{eqnarray}
e_{ab}^+(\hat\Omega)
&=&
\hat l_a\hat l_b-\hat m_a\hat m_b\,,
\\
e_{ab}^\times(\hat\Omega)
&=&
\hat l_a\hat m_b+\hat m_a\hat l_b\,.
\end{eqnarray}
Note that there is a rotational
degree of freedom in the definition of polarization
tensors as one is free to rotate $\hat l$ and
$\hat m$ by an angle $\psi$ 
in the plane orthogonal to $\hat\Omega$.
For a gravitational-wave source with a
symmetry axis, such as an inspiralling binary, 
the angle 
$\psi$ can be interpreted as the polarization
angle of the source.
However, as we will assume that the stochastic
background is unpolarized, there is no loss
of generality in taking $\psi=0$, so that the
polarization tensors have the form given above.

The Fourier coefficients $h_A(f,\hat{\Omega})$ are
complex functions that satisfy
$h_A(-f,\hat{\Omega})=h_A^*(f,\hat\Omega)$, since
$h_{ab}(t,\vec x)$ is real.
For a stochastic gravitational-wave background these 
coefficients are {\em random} fields whose expectation values 
define the statistical properties of the background.
Without loss of generality we can assume that the
fields have zero mean:
\begin{equation}
\langle h_A(f,\hat \Omega)\rangle = 0
\,.
\end{equation}
We will also assume that the background is
unpolarized, Gaussian, and stationary, but 
allow for an anisotropic distribution.
The most general form of the quadratic expectation
value satisfying these requirements is
\begin{equation}
\langle h^*_A(f,\hat\Omega) h_{A'}(f',\hat\Omega')\rangle
=\frac{1}{4}{\cal P}(f,\hat\Omega)\,
\delta(f-f')
\delta_{AA'}
\delta(\hat\Omega,\hat\Omega')\,,
\label{e:hAhA'}
\end{equation}
where ${\cal P}(f,\hat\Omega)$ specifies both the 
spectral and angular distribution of the
background.
The factor of $1/4$ has been included so that
for an isotropic background
$H(f)\equiv {\cal P}(|f|,\hat\Omega)$ 
is the {\em one-sided} strain power, 
when summed over {\em both} polarizations.
Given the above definitions, 
it follows that
\begin{equation}
\Omega_{\rm gw}(f)
\equiv
\frac{f}{\rho_{\rm c}}
\frac{d\rho_{\rm gw}}{df}
=\frac{2\pi^2}{3H_0^2}{f^3}
\int_{S^2}d\hat\Omega\>
{\cal P}(f,\hat\Omega)
\,,
\label{e:OmegaGW}
\end{equation}
where $d\rho_{\rm gw}$ is the energy density contained
in the frequency interval $df$.
Here
$H_0$ is Hubble's constant, and
$\rho_c\equiv 3 c^2 H_0^2/8\pi G$ is the critical energy
density needed to close the universe.
(To prove Eq.~(\ref{e:OmegaGW}), one should write 
$\rho_{\rm gw}$ in terms of an expectation
value of the product of the time derivatives
of the metric perturbations 
$h_{ab}(t,\vec x)$,
and then expand the metric perturbations in 
terms of the plane wave components as in 
Eq.~(\ref{e:hab}), using Eq.~(\ref{e:hAhA'}) 
to evaluate the expectation value; 
see, e.g., \cite{Allen:1999, allen-ottewill}.)
Thus, the energy density in a stochastic 
gravitational-wave background has
contributions from all parts of the sky 
as encoded in the all-sky integral of
$\mathcal{P}(f,\hat\Omega)$.

In what follows, we will assume 
that ${\cal P}(f,\hat\Omega)$ can be
factorized into a product of two functions
\begin{equation}
{\cal P}(f,\hat\Omega)
=
{\cal P}(\hat \Omega)\bar{H}(f)
\,,
\label{e:PH}
\end{equation}
where $\bar{H}(f)$ is a dimensionless
function of frequency, normalized so that
$\bar H(f_R)=1$, where $f_R$ is a 
reference frequency,
typically taken to equal $\unit[100]{Hz}$
(a frequency in LIGO's most sensitive band).
$\mathcal{P}(\hat \Omega)$ specifies the
angular distribution of gravitational-wave
power, and $\bar H(f)$ its spectral
shape.
This factorization does not amount to a loss of 
generality if one restricts attention
to small enough frequency bands.
For our analysis, we will assume that 
\begin{equation}
\bar {H}(f)=(f/f_R)^\beta
\label{e:Hbar}
\,,
\end{equation}
where $\beta$ is a power-law index which we
fix (for example, 
$\beta=0$ for constant strain power).
Using Eqs.~(\ref{e:OmegaGW}) and (\ref{e:PH}),
one can show that this assumption for
$\bar H$ is consistent with
\begin{equation}
\Omega_{\rm gw}(f)=\Omega_R(f/f_R)^{3+\beta}\,,
\label{e:OmegaR}
\end{equation}
where $\Omega_R$ is the fractional energy density
in gravitational waves evaluated at the reference
frequency $f_R$.

The angular distribution function 
${\cal P}(\hat\Omega)$ can be expanded in terms 
of a set of basis functions on the two-sphere
according to
\begin{equation}
{\cal P}(\hat\Omega)=
{\cal P}_\alpha 
{\bf e}_\alpha(\hat\Omega)
\,, \label{eq:defPAlpha}
\end{equation}
where summation (or integration) over 
$\alpha$ is understood, and
\begin{eqnarray}
{\cal P}_{\alpha} 
&=&
\int_{S^{2}} d\hat{\Omega}\>
{\cal P}(\hat{\Omega}){\bf e}_{\alpha}^{*}(\hat{\Omega})
\,,
\\
\delta_{\alpha\beta} 
&=&
\int_{S^{2}} d\hat{\Omega}\>
{\bf e}_{\alpha}^{*}(\hat{\Omega}){\bf e}_{\beta}(\hat{\Omega})
\,.
\end{eqnarray}
The choice of basis, in principle, should not
affect the physical search results. However,
in practice, such a choice can bear on computational costs
of a search and also on the systematic
errors affecting observations results, e.g.,
arising from the truncation order of the 
spherical-harmonic basis. For these reasons, 
we expect that while searching for 
gravitational-wave point sources, a decomposition with 
respect to a pixel basis
\begin{equation}
{\cal P}(\hat\Omega)
=
{\cal P}_{\hat\Omega'}
\delta(\hat\Omega,\hat\Omega')
\label{e:POmega}
\end{equation}
%
is the natural choice.
For a diffuse background, e.g., dominated
by a dipole or quadrupolar distribution, a
spherical harmonic decomposition may be the better choice:
\begin{eqnarray}
{\cal P}(\hat\Omega)
&=&
{\cal P}_{lm}Y_{lm}(\hat\Omega)
\,,
\\
{\cal P}_{lm} 
&=&
\int_{S^2}d\hat\Omega\,
{\cal P}(\hat\Omega)Y^*_{lm}(\hat\Omega)
\,,
\label{e:plm}
\end{eqnarray}
where the second equality follows from our
normalization convention for the $Y_{lm}$
(see Appendix~\ref{s:Ylms}).
Note that the pixel basis coefficients, ${\cal P}_{\Omega^\prime}$, 
have units of $\unit[]{strain^2/Hz}$ 
whereas the coefficients in the spherical harmonics basis, 
${\cal P}_{lm}$, have units of $\unit[]{strain^2/Hz/rad}$.
This normalization convention also implies 
\begin{equation}
\Omega_R=
\frac{2\pi^2}{3 H_0^2}
{f_R^3}
\,
\sqrt{4 \pi}
{\cal P}_{00}
\,.
\label{e:OmegaR_P00}
\end{equation}
Note that only the monopole moment ${\cal P}_{00}$
contributes to $\Omega_R$ (and hence to $\Omega_{\rm gw}(f)$)
as all higher-order multipole moments give zero when 
integrated over the sky.

\subsection{Overlap factor}

We will denote the time-series output of two 
detectors $I=1,2$ by
\begin{equation}
s_I(t)=h_I(t) + n_I(t)
\,,
\end{equation}
where $n_I(t)$ is the detector noise and
$h_I(t)$ is its response to a gravitational-wave 
background: 
\begin{equation}
h_I(t)=
\int_{-\infty}^\infty df
\int_{S^2}d\hat\Omega\,
h_A(f,\hat\Omega)
F_I^A(\hat\Omega,t)\,
e^{i 2\pi f(t-\hat{\Omega}\cdot \vec x_I(t)/c)}\,.
\label{e:hI(t)}
\end{equation}
Here
\begin{equation}
F_I^A(\hat\Omega,t)
=d_I^{ab}(t)e_{ab}^A(\hat\Omega)
\label{e:F_I}
\end{equation}
is the detector response function, which encodes the
directional sensitivity of detector $I$ to a 
plane-polarized gravitational wave propagating in direction
$\hat\Omega$ and $\vec{x}_I$ specifies the location of interferometer $I$.
(The absolute value $|F_I^A(\hat\Omega,t)|$ plotted as
function of direction $\hat\Omega$ is called the detector
{\em antenna pattern}.) 
The detector tensor is
\begin{equation}
d_I^{ab}(t)=
\frac{1}{2}
\left[\hat X_I^a(t)\hat X_I^b(t)-\hat Y_I^a(t)\hat Y_I^b(t)\right]
\,,
\label{e:d_I}
\end{equation}
where
$\hat X_I(t)$, $\hat Y_I(t)$ are unit vectors pointing
along the interferometer arms 
for detector $I$.
The vectors $\vec x_I(t)$, $\hat X_I(t)$, and $\hat Y_I(t)$ are 
all time-dependent due to the Earth's rotation.
(We are using {\em equatorial} coordinates, with the 
spatial origin at the center of the Earth, $\hat z$-axis 
pointing along the Earth's rotation axis, and 
$\hat x$-axis pointing in the
direction of the vernal equinox.)

Given a time-series $s_I(t)$, we define its {\em short-term 
Fourier transform} $\tilde s_I(f,t)$ by
\begin{equation}
\tilde s_I(f,t)
\equiv
\int_{t-\tau/2}^{t+\tau/2}dt'\,
e^{-i2\pi f t'} s_I(t')
\,,
\label{e:SFT}
\end{equation}
where $\tau$ is much greater than the light-travel time
between any pair of detectors, but is small enough that the
the detector response function $F_I^A(\hat\Omega,t)$ and 
detector location $\vec x_I(t)$ do not vary significantly 
with time 
over the interval $[t-\tau/2,t+\tau/2]$.
Typical values of $\tau$ are from a few tens of seconds
to a few hundred seconds.
The cross-correlation between the output of the two detectors 
is then defined in terms of these short Fourier transforms as
\begin{equation}
C(f,t) \equiv 
\frac{2}{\tau}\,
\tilde s_1^*(f,t) \tilde s_2(f,t)
\,.
\label{e:C(f,t)}
\end{equation}
The factor of 2 is a convention consistent with the definition of
one-sided power spectra, so that the total cross-power for a 
particular time $t$ is given by integrating $C(f,t)$ over positive 
frequencies.
These cross-spectra are the starting point for the maximum-likelihood
analysis described in the following section.

If the noise at the two detectors is uncorrelated---a reasonable 
assumption for spatially-separated detectors---then it follows 
that the expectation value of the cross-spectra depends only on 
the gravitational-wave signal components
\begin{equation}
\langle C(f,t)\rangle
=
\frac{2}{\tau}\,
\langle \tilde h_1^*(f,t) \tilde h_2(f,t)\rangle\,.
\label{e:<C(f,t)>}
\end{equation}
Using
%
%
Eqs.~(\ref{e:hAhA'}), (\ref{e:PH}), and the
short-term Fourier transform of (\ref{e:hI(t)}),
one can then show that
\begin{equation}
\langle C(f,t)\rangle 
=
\bar H(f)
\int_{S^2} 
d\hat\Omega\,
\gamma(\hat\Omega,f,t) 
{\cal P}(\hat\Omega)
\,,
\label{e2:<C(f,t)>}
\end{equation}
where
\begin{equation}
\gamma({\hat\Omega},f,t)
=
\frac{1}{2}
F_1^A(\hat\Omega,t)F_2^A(\hat\Omega,t) 
e^{i2\pi f\hat\Omega\cdot(\vec x_1(t)-\vec x_2(t))/c}
\,.
\label{e:gamma(Omega,f,t)}
\end{equation}
The function $\gamma({\hat\Omega},f,t)$ is a 
geometric factor that takes into account the 
separation and relative orientation of the two detectors 
(see e.g.,~\cite{Finn}).
For an isotropic background,
$\langle C(f,t)\rangle \propto \bar H(f)\gamma(f)$, where
\begin{equation}
\gamma(f)
\equiv
\frac{5}{8\pi}
\int_{S^2}d\hat\Omega\,
F_1^A(\hat\Omega,t)F_2^A(\hat\Omega,t) 
e^{i2\pi f\hat\Omega\cdot(\vec x_1(t)-\vec x_2(t))/c}
\end{equation}
is the standard 
{\em overlap reduction function}~\cite{christensen:1992,flanagan:1993}.
The factor of ${5/8\pi}$ is a normalization constant 
chosen so that 
$\gamma(f)=1$ for all frequencies for a pair of
coincident and coaligned interferometers with 90-degree
opening angle between the interferometer arms.
Note that $\gamma(f)$ is time-independent due to the 
all-sky integration.

For subsequent analysis, it will be convenient to rewrite 
the right-hand side of Eq.~(\ref{e2:<C(f,t)>}) 
in terms of an integral over the components of
${\cal P}(\hat\Omega)$ and $\gamma(\hat\Omega,f,t)$ 
with respect to  a set 
of basis functions on the two-sphere
\begin{equation}
\langle C(f,t)\rangle 
= 
\bar H(f)
\gamma_\alpha(f,t)
{\cal P}_\alpha
\,,
\end{equation}
where
\begin{eqnarray}
\gamma_{\alpha}(f, t)
&=&
\int_{S^{2}}
d\hat{\Omega}\>
\gamma(\hat{\Omega}, f, t){\bf e}_{\alpha}(\hat{\Omega})\,,
\label{e:gamma_alpha}
\\
\gamma(\hat{\Omega}, f, t)
&=&
\gamma_{\alpha}(f, t){\bf e}_{\alpha}^{*}(\hat{\Omega})
\,.
\end{eqnarray}
Therefore, in the pixel basis, 
$\alpha\leftrightarrow\hat\Omega'$, one has
\begin{equation}
\gamma(\hat\Omega,f,t)
=
\gamma_{\hat\Omega'}(f,t)\delta(\hat\Omega,\hat\Omega')
\label{e:gammaOmega}
\,,
\end{equation}
%
and in the spherical harmonics basis,
$\alpha\leftrightarrow lm$, one gets
\begin{eqnarray}
\gamma(\hat\Omega,f,t)
&=&
\gamma_{lm}(f,t)
Y^*_{lm}(\hat\Omega)
\,,
\\
\gamma_{lm}(f,t)
&=&
\int_{S^2} d\hat\Omega\,
\gamma(\hat\Omega,f,t)
Y_{lm}(\hat\Omega)
\,.
\label{e:gammalm}
\end{eqnarray}
Note that the above definition of $\gamma_{lm}(f,t)$ 
differs from that (\ref{e:plm}) 
of ${\cal P}_{lm}$ by a complex conjugation, 
but it agrees with the
convention used in \cite{allen-ottewill}.
The time-dependence of $\gamma_{lm}(f,t)$ is particularly
simple:
\begin{equation}
\gamma_{lm}(f,t)=
\gamma_{lm}(f,0)\,
\exp
\left(
im 2\pi \frac{t_{\rm sidereal}}{\unit[1]{sidereal~day}}
\right)\,.
\label{e:gamma_lm(t)}
\end{equation}
In addition, for an isotropic background, the 
above definitions imply
$\gamma(f)=(5/\sqrt{4\pi})\gamma_{00}(f,t)$ for any $t$.

\section{MAXIMUM-LIKELIHOOD ANALYSIS}
\label{s:MLestimation}

In this section, we derive the maximum-likelihood estimators
of the angular distribution ${\cal P}_\alpha$ of the 
stochastic gravitational-wave power.
The analysis given below also makes clear the relationship 
between the 
covariance matrix of these estimators and the beam pattern 
matrix for the cross-correlation measurements.
Since maximizing the likelihood is equivalent to minimizing the
squared deviation of the estimators away from their expected
values, the estimators so obtained are {\em optimal} in the 
sense of maximizing the expected signal-to-noise ratio of the
estimators.
Hence, the likelihood analysis reproduces the standard results
of optimal filtering for the isotropic 
(e.g., \cite{S1Isotropic,S3Isotropic,S4Isotropic})
and radiometer 
\cite{mitra-et-al,radiometer} 
analyses without explicitly introducing a filter function $Q$ 
in the construction of a statistic.

\subsection{Maximum-likelihood estimators}
\label{s:MLestimators}

The maximum likelihood analysis for an 
anisotropic stochastic background takes as its
fundamental data vector the cross-spectra 
\begin{equation}
C_{ft}
\equiv
C(f,t)
=
\frac{2}{\tau}
\tilde s^*_1(f,t)\tilde s_2(f,t)
\end{equation}
evaluated at a set of discrete times $t$ and
discrete (positive and negative) frequencies $f$.
As shown in the previous subsection, 
the expectation values 
of the cross-spectra are given by
\begin{equation}
\langle C_{ft}\rangle
=
\bar H(f)
\gamma_\alpha(f,t)
{\cal P}_\alpha
\,.
\label{e:chgp}
\end{equation}
The covariance matrix is given by
\begin{eqnarray}
N_{ft,f't'}
&=&
\langle C_{ft}C^*_{f't'}\rangle-
\langle C_{ft}\rangle
\langle C^*_{f't'}\rangle
\\
&\approx&
\delta_{tt'}
\delta_{ff'}
P_{1}(f,t) P_{2}(f,t)
\label{e:N}
\end{eqnarray}
where $P_I(f,t)$, $I=1,2$ are the one-sided
power spectra of the detector output for 
time segment $t$, which satisfy
\begin{eqnarray}
\langle \tilde s^*_I(f,t)\tilde s_I(f',t')\rangle
&=&
\frac{\tau}{2}
\delta_{tt'}
\delta_{ff'}
P_I(f,t)
\label{e:P_I(f,t)}
\\
&\approx&
\langle \tilde n^*_I(f,t)\tilde n_I(f',t')\rangle
\label{e:P_nI(f,t)}
\,.
\end{eqnarray}
We have assumed that there is no cross-correlated noise, 
and that 
the cross-correlated and auto-correlated
gravitational-wave signal power are much less 
than the detector noise power to obtain the 
approximate relations
Eqs.~(\ref{e:N}) and (\ref{e:P_nI(f,t)}).

Treating $P_I(f,t)$ and the gravitational-wave spectral shape 
$\bar H(f)$ as known quantities, the likelihood function is then
\begin{equation}
p(\{C_{ft}\}|\{{\cal P}_\alpha\})
\propto
\exp\bigg[-\frac{1}{2}
\chi^2({\cal P})\bigg]
\,,
\label{e:likelihood}
\end{equation}
where 
\begin{equation}
\chi^2({\cal P})
\equiv
\sum_{tft'f'}
(C^*_{ft}-\langle C^*_{ft}\rangle)
N^{-1}_{ft,f't'}
(C_{f't'}-\langle C_{f't'}\rangle)\,.
\end{equation}
Using Eqs.~(\ref{e:chgp}) and (\ref{e:N}),
we have
\begin{align}
\chi^2({\cal P})
=&\sum_t \sum_f
(C^*(f,t)-\bar H(f)\gamma_\alpha^*(f,t){\cal P}^*_\alpha)
\nonumber
\\
&\frac{1}{P_1(f,t) P_2(f,t)}
(C(f,t)-\bar H(f)\gamma_\beta(f,t){\cal P}_\beta)
\,.
\end{align}
Since maximizing the likelihood with 
respect to ${\cal P}_\alpha$ 
is equivalent to 
minimizing chi-squared,
one can show that the 
maximum likelihood estimators for the ${\cal P}_\alpha$ 
are given by
\begin{equation}
\hat{\cal P}_\alpha 
= 
(\Gamma^{-1})_{\alpha\beta}\, X_\beta
\,,
\label{e:Phat}
\end{equation}
where 
%
\begin{eqnarray}
X_{\beta}
=
\sum_t 
\sum_f
\gamma_{\beta}^*(f,t)\,
\frac{\bar H(f)}{P_1(f,t) P_2(f,t)}\,
C(f,t)
\,,
\label{e:X}
\\
\Gamma_{\alpha\beta}
=
\sum_t
\sum_f
\gamma_{\alpha}^*(f,t)\,
\frac{\bar {H}^2(f)}{P_1(f,t) P_2(f,t)}\,
\gamma_{\beta}(f,t)
\,.
\label{e:Gamma}
\end{eqnarray}
(We will adress the invertability of 
$\Gamma_{\alpha\beta}$ in section \ref{s:data_analysis}.)
Note that the standard estimator of the strength
of an isotropic stochastic background~\cite{S1Isotropic,S3Isotropic,S4Isotropic}
\begin{equation}
\hat\Omega_{\rm gw} = 
\left(\sum_t\frac{1}{\sigma_t^2}\right)^{-1}
\sum_t \frac{Y_t}{\sigma_t^2}
\end{equation}
has the same form as the above, 
with $\sum_t Y_t/\sigma_t^2$ playing the role of $X_\beta$
and $\sum_t 1/\sigma_t^2$ the role of $\Gamma_{\alpha\beta}$.

For later reference, we note that  
the minimum value of chi-squared is
$\chi^2_\text{min}=\chi^2(\hat{\cal P})$, which
can be written explicitly as
\begin{align}
\chi^2_\text{min}
&= \sum_t\sum_f \frac{|C(f,t)|^2}{P_1(f,t)P_2(f,t)} 
\nonumber
\\
&\qquad
- \hat{\cal P}^*_\alpha X_\alpha 
- X^*_\beta \hat{\cal P}_\beta 
+ \hat{\cal P}^*_\alpha\Gamma_{\alpha\beta}\hat{\cal P}_\beta
\,.
\label{e:chi2min}
\end{align}
Also, in analogy with what is done for cosmic microwave 
background experiments such as WMAP \cite{WMAP}, we can 
construct estimators of the angular ``power" spectra 
\begin{equation}
C_l = \frac{1}{2l+1}\sum_{m=-l}^l |{\cal P}_{lm}|^2
\end{equation}
by simply replacing ${\cal P}_{lm}$ with the estimators
$\hat{\cal P}_\alpha$ evaluated in the spherical harmonics
basis---i.e.,
\begin{equation}
\hat C_l = \frac{1}{2l+1}\sum_{m=-l}^l |\hat{{\cal P}}_{lm}|^2
\,.
\label{e:hatCl}
\end{equation}
Note that the $\hat C_l$ defined above are actually estimators 
of the {\em squared} angular power (since $\hat{\cal P}_{lm}$
already has units of power), unlike the cosmic microwave 
background data for 
which the $\hat C_l$ really do have units of power.
Also, we will see in the next subsection that these 
estimators are {\em biased}.
Unbiased estimators of the $C_l$ are given in equation~(\ref{e:hatCl'}).

\subsection{Error estimates}

If the spectral shape $\bar H(f)$ that we assumed for
the signal model exactly matches that of the 
observed background, it is fairly easy to show that
the estimators $\hat{\cal P}_\alpha$ constructed 
above provide unbiased estimates of the angular
distribution of gravitational-wave power:
\begin{equation}
\langle \hat{\cal P}_\alpha\rangle
=
{\cal P}_\alpha
\,.
\end{equation}
This follows immediately from the fact that
\begin{equation}
\langle X_\alpha\rangle
=
\Gamma_{\alpha\beta}\,
{\cal P}_\beta
\,
\label{e:<X>},
\end{equation}
which in turn shows that the $X_\alpha$ are the 
components of the so-called `dirty' map---i.e.,
$X_\alpha$ represent 
the gravitational-wave power on the sky 
as seen through the beam matrix of the two detectors,
$\Gamma_{\alpha\beta}$.
Equation (\ref{e:Phat}) shows that
by inverting $\Gamma_{\alpha\beta}$, 
one obtains 
the components of the `clean' map, $\hat{\cal P}_\alpha$.
The process of going from the dirty map to the clean
map is an example of {\em deconvolution}.
 
In addition, one can show in the weak-signal 
approximation that 
\begin{eqnarray}
&&
\langle X_\alpha X^*_\beta\rangle-
\langle X_\alpha\rangle
\langle X^*_\beta\rangle
\approx
\Gamma_{\alpha\beta}\,,
\label{e:<XX>}
\\
&&
\langle \hat{\cal P}_\alpha \hat{\cal P}^*_\beta\rangle -
\langle \hat{\cal P}_\alpha\rangle
\langle \hat{\cal P}^*_\beta\rangle 
\approx
(\Gamma^{-1})_{\alpha\beta}
\,.
\label{e:<PP>}
\end{eqnarray}
Thus, $\Gamma_{\alpha\beta}$ is the covariance matrix
of the dirty map $X_\alpha$, and 
$(\Gamma^{-1})_{\alpha\beta}$ is the covariance matrix
of the clean map $\hat{\cal P}_\alpha$.
A matrix like $\Gamma_{\alpha\beta}$, whose inverse 
is the covariance matrix of the signal parameters, is
often called a {\em Fisher} information matrix.
An alternative definition of $\Gamma_{\alpha\beta}$,
illustrating its connection to the likelihood function,
is
\begin{equation}
\Gamma_{\alpha\beta}
=
-
\bigg\langle
\frac{\partial^2 \ln p(\{C_{ft}\}|\{{\cal P}_\alpha)\}}
{\partial {\cal P}^*_\alpha\partial {\cal P}_\beta}
\bigg\rangle
\,.
\label{e:FisherPartial}
\end{equation}
As is evident from the above expression, if one 
has several independent measurements (so that the 
combined likelihood is just a product of individual 
likelihoods), the 
$\Gamma_{\alpha\beta}$ matrices simply add.  

Finally,
using the above expressions for the expectation
value and covariances of the $\hat{\cal P}_\alpha$,
one can show that
\begin{eqnarray}
\langle
\hat C_l
\rangle
&\approx& 
C_l 
+\frac{1}{2l+1}\sum_m
(\Gamma^{-1})_{lm,lm}
\,,
\\
\langle \hat{C}_l^2 \rangle
-
\langle \hat{C}_l \rangle^2
&\approx&
\frac{2}{(2l+1)^2}
\sum_{m,m'}
|(\Gamma^{-1})_{lm,lm'}|^2
\,.
\end{eqnarray}
Note, in particular, that the estimators $\hat C_l$ are biased.
Unbiased estimators of the $C_l$ are given by
\begin{equation}
\hat C_l'
= \hat C_l-\frac{1}{2l+1}\sum_m
(\Gamma^{-1})_{lm,lm}\,.
\label{e:hatCl'}
\end{equation}

\subsection{Decomposition in terms of pixel basis or 
spherical harmonics}

The analysis presented above has been written in terms
of the components 
${\cal P}_\alpha$ and $\gamma_\alpha(f,t)$
of 
${\cal P}(\hat\Omega)$ and $\gamma(\hat\Omega,f,t)$ 
with respect to an arbitrary
set of basis functions on the two-sphere.
For most purposes, we will be interested in the 
components with respect to only two bases:
The pixel basis, for which 
$\alpha\leftrightarrow\hat\Omega$ and
${\cal P}_{\hat\Omega}$ and
$\gamma_{\hat\Omega}(f,t)$ are given by
(\ref{e:POmega}) and (\ref{e:gammaOmega}), and
the spherical harmonics basis, for which 
$\alpha\leftrightarrow lm$ and
${\cal P}_{lm}$ and $\gamma_{lm}(f,t)$ are given
by (\ref{e:plm}) and (\ref{e:gammalm}).
Each basis has its own set of advantages and
disadvantages, which we briefly describe below.

In the pixel basis,
$\hat{\cal P}_{\hat\Omega}$ is an estimate of 
the true gravitational-wave power ${\cal P}_{\hat\Omega}$
coming from direction
$\hat\Omega$.
It is a real quantity and should be non-negative.
The quantity
$X_{\hat\Omega}$, on the other hand, is the 
power coming from direction $\hat\Omega$ 
{\em as seen by the detector}.
It includes gravitational-wave power from other 
directions on the sky due to the
finite acceptance of the beam pattern
function, as well as from instrumental noise.
The matrix $\Gamma_{\hat\Omega\hat\Omega'}$
connects the two via (\ref{e:<X>}), and
can be directly interpreted as a 
{\em point spread function}.
It specifies
how a point source at $\hat\Omega$
is spread to other points $\hat\Omega'$ by 
the response of a pair of detectors.

In the spherical harmonics basis, the
$\hat{\cal P}_{lm}$ are estimates of the true
{\em multipole moments} ${\cal P}_{lm}$ 
of the gravitational-wave power on the sky.
The matrix 
$\Gamma_{lm,l'm'}$, is no longer directly
interpretable in terms of a point spread function,
but it plays an analogous role as the 
inverse matrix $(\Gamma^{-1})_{lm,l'm}$
specifies the correlations between
the various multipole moment estimates.

In addition, the Fisher matrix $\Gamma_{\alpha\beta}$ has two symmetries:
parity (see Eq.~\ref{eqYmo}) and rotational symmetry around the z-axis.
Since spherical harmonics respect these symmetries,
this leads to some simplifications.
Parity is an exact symmetry, because the only difference
between gravitational-wave 
signals coming from antipodes is an opposite
sign of the time shift between detectors. Therefore the
detector noise, as expressed by the Fisher matrix, is identical
for antipodes. This implies that $\Gamma_{lm,l'm'}=0$
for all odd $l-l'$ (almost half of the matrix elements).
Z-axis rotational symmetry is broken by
daily variations in detector sensitivity, but still implies
that $\Gamma_{lm,l'm'} \approx 0$ for $m \neq m'$, i.e.,
$\Gamma_{lm,l'm'}$ is a block-diagonally dominant matrix.
The pixel basis has no such symmetry.

Furthermore, in the spherical harmonics basis 
it is simple to specify a resolution cut-off by only
allowing $l \leq l_{\rm max}$. This avoids over-sampling and
reduces the number of required basis vectors.
Also, since extending this cut-off to a larger $l_{\rm max}$
does not affect the original basis vectors, it is
straightforward to run the analysis with a higher resolution,
and later do the matrix inversion at a lower resolution.

Finally, the computationally dominant part of the analysis is
the calculation of the Fisher matrix.
Since the Fisher matrix has $N^2$ elements, with $N$ the
number of basis vectors, working in the 
spherical harmonics basis
makes the analysis significantly more efficient. 
And the mentioned symmetries
help to reduce the computational load even more.

\section{IMPLEMENTATION AND ANALYSIS DETAILS}
\label{s:data_analysis}

As shown in Sec.~\ref{s:MLestimators}, the maximum-likelihood
estimators of the angular distribution of power in an anisotropic
gravitational-wave background are given by
\begin{equation}
\hat{\cal P}_\alpha 
= 
(\Gamma^{-1})_{\alpha\beta}\, X_\beta
\,,
\label{e:Phat2}
\end{equation}
where $X_\beta$ are the components of the `dirty' map (\ref{e:X}),
and $(\Gamma^{-1})_{\alpha\beta}$ are the components of the inverse of
the beam pattern matrix $\Gamma_{\alpha\beta}$
(\ref{e:Gamma}).  In this section, we describe:
(i) some of the
implementation details related to the calculation of $X_\beta$ and
$\Gamma_{\alpha\beta}$, 
(ii) a method for regularizing the
inversion of $\Gamma_{\alpha\beta}$, 
(iii) how to extend the single baseline analysis to a network of
detectors, and 
(iv) a Bayesian model selection scheme for determining
$l_{\rm max}$ for the spherical harmonic decomposition.
For concreteness we consider a network of detectors 
consisting of the LIGO interferometers H1 and L1
and the Virgo interferometer, V1.

\subsection{Calculating $X_\beta$ and $\Gamma_{\alpha\beta}$}

The components of the `dirty' map $X_\beta$ and the beam pattern 
matrix $\Gamma_{\alpha\beta}$ are given by
\begin{eqnarray}
X_{\beta}
=
\sum_t 
\sum_f
\gamma_{\beta}^*(f,t)\,
\frac{\bar H(f)}{P_1(f,t) P_2(f,t)}\,
C(f,t)
\,,
\label{e:X2}
\\
\Gamma_{\alpha\beta}
=
\sum_t
\sum_f
\gamma_{\alpha}^*(f,t)\,
\frac{\bar {H}^2(f)}{P_1(f,t) P_2(f,t)}\,
\gamma_{\beta}(f,t)
\,.
\label{e:Gamma2}
\end{eqnarray}
These are the fundamental data products of this analysis 
from which $(\Gamma^{-1})_{\alpha\beta}$ 
and $\hat{\cal P}_\alpha$ are then calculated (\ref{e:Phat2}).
Although the various quantities entering $X_\beta$ and 
$\Gamma_{\alpha\beta}$ have already been defined in the previous 
two sections, 
we describe here in more detail how they are calculated in practice.

(i) $\bar H(f) = (f/f_R)^\beta$ is the assumed spectral shape of the 
gravitational-wave background (\ref{e:Hbar}).
This is an input to the data analysis pipeline which we fix at the 
start of the analysis.
The parameters $f_R$ and $\beta$ are the reference frequency and 
spectral index for the (assumed) power-law behavior of the 
gravitational-wave spectrum.
For the analyses described later in this paper, we choose
$f_R=100\ {\rm Hz}$ and $\beta=0$, corresponding to constant 
strain power.
Other values of $f_R$ and $\beta$ are, of course, possible.
For example, $\beta=-3$ corresponds to constant fractional
energy density
$\Omega_{\rm gw}(f) = {\rm const}$, which follows from Eq.~(\ref{e:OmegaR}).

(ii) $C(f,t)$ are the cross-spectra of the data,
calculated as a product of the short-term Fourier transforms 
of the time-series output of the two detectors,
cf.~Eqs.~(\ref{e:C(f,t)}), (\ref{e:SFT}).
The time-series data are first downsampled to a few
kilohertz (from 16384~Hz to 2048~Hz for LIGO;
from 10000~Hz to 2000~Hz for Virgo),
high-pass filtered above 40~Hz (to reduce contamination
from low-frequency seismic noise), and then windowed 
(to avoid spectral leakage of strong instrumental lines),
before being discrete-Fourier-transformed to the frequency
domain.
As $\tau$ is typically of order 100~s
($\tau=60~{\rm s}$ for the simulations that
we will describe in Sec.~\ref{s:simulations}),
the frequency resolution of
$\tilde s_I(f,t)$ and $C(f,t)$ is of order
$1/\tau = 0.01\ {\rm Hz}$, which is much finer than what
is needed for the other frequency-series
$\bar H(f)$, $P_I(f,t)$, and $\gamma_\alpha(f,t)$,
which are typically taken to have a frequency
resolution $\Delta f=0.25\ {\rm Hz}$.
Hence, to match $\Delta f$, we average together 
several frequency bins of $C(f,t)$.
This averaging or ``coarse graining" has also been used for 
previous stochastic searches, see e.g. \cite{S1Isotropic}.
It is a technique used to avoid 
unnecessary frequency resolution, especially in $P_I(f,t)$ and 
$\gamma_\alpha(f,t)$.

(iii) $P_I(f,t)$ are the power spectra 
associated with the individual ($I=1,2$) 
detector outputs (\ref{e:P_I(f,t)}).
We use Welch's modified periodogram method to estimate the
power spectra, averaging together periodograms from 4-sec long, 
50\% overlapping, Hann-windowed data, which are taken from the 
two time segments immediately preceding and following---but not 
including---the analyis segment.
(The 4-sec data stretches corresponds to the $\Delta f=0.25$~Hz 
frequency resolution mentioned earlier.)
This technique greatly reduces a bias that would otherwise result from 
non-zero covariance between $P_I(f,t)$ and $C(f,t)$.

For an actual analysis of real data, one needs to consider the possibility of short-term variations in the detector noise that are not consistent between the analysis segment and the two neighboring segments from which the $P_I(f,t)$ were estimated.
For the simulations that we will describe in Sec.~\ref{s:simulations}, the data were stationary, so no consistency cut needed to be applied.

(iv) $\gamma_\alpha(f,t)$
are the components of the overlap factor 
$\gamma(\hat\Omega,f,t)$ defined by 
Eqs.~(\ref{e:gamma(Omega,f,t)}), (\ref{e:gammaOmega}), (\ref{e:gammalm}).
These are geometric factors that encode the relative 
separation and orientation of the two detectors,
as specified by the detector response functions
$F_I^A(\hat\Omega,t)$ 
(\ref{e:F_I}),
(\ref{e:d_I}).  
In the pixel basis, the components $\gamma_{\hat\Omega'}(f,t)$
can be efficiently calculated by using one Fast Fourier Transform,
and reading out the resulting cross-correlation time series at the
time shift corresponding to each pixel \cite{Ballmer:2006,radiometer}.
In the spherical harmonics basis, the components 
$\gamma_{lm}(f,t)$ can be 
efficiently calculated using analytic expressions 
derived in \cite{allen-ottewill}.
In particular, the authors in \cite{allen-ottewill}
show that for sidereal 
time $t=0$, one can write 
$\gamma_{lm}(f,0)$ as a simple linear combination of 
spherical bessel functions $j_n(x)/x^n$ (for $l$ even)
or $j_n(x)/x^{n-1}$ (for $l$ odd), where $x$ depends 
on the relative separation of the detectors 
$x=2\pi f|\vec x_1-\vec x_2|/c$.
The coefficients of the linear combinations are
complex numbers that depend 
on the relative orientation of the detectors.
Explicit expressions for a few of the 
$\gamma_{lm}(x)\equiv \gamma_{lm}(f,0)$
for the LIGO Hanford-Livingston pair are given below:
\begin{align}
\gamma_{00}(x) 
&=
-0.0766 j_0(x) 
-2.1528 \frac{j_1(x)}{x}
\nonumber\\
&\qquad\qquad + 2.4407 \frac{j_2(x)}{x^2}
\,,
\\
\gamma_{10}(x) 
&=
-0.0608i\, j_1(x) 
-2.6982i\, \frac{j_2(x)}{x}
\nonumber\\
&\qquad\qquad +7.7217i\,\frac{j_3(x)}{x^2}
\,,
\\
\gamma_{11}(x) 
&=
-(0.0519+0.0652i)\, j_1(x) 
\nonumber\\
&\qquad\qquad 
-(1.8622+1.0516i)\, \frac{j_2(x)}{x}
\nonumber\\
&\qquad\qquad 
+(4.0106 - 2.4936i)\,\frac{j_3(x)}{x^2}
\,.
\end{align}
(Note that the numerical coefficients above do 
not agree with those in \cite{allen-ottewill}, due to 
an overall normalisation by $4\pi/5$
and phase factor $e^{im\phi}$, where $\phi=-38.52^\circ$
is the angle between the separation vector 
between the LIGO Hanford and Livingston detectors
and the Greenwich meridian.)
For arbitrary sidereal times $t$, one uses 
Eq.~(\ref{e:gamma_lm(t)}), which follows from the 
$e^{im\phi}$ dependence of the spherical harmonics
$Y_{lm}(\hat\Omega)=Y_{lm}(\theta,\phi)$.
Here $(\theta,\phi)$ are related to the equatorial 
coordinates (ra,dec) via 
$\theta =\pi/2- \pi\,({\rm dec}/180^\circ)$
and
$\phi = \pi\, ({\rm ra}/12\ {\rm hr})$.

\subsection{Deconvolution and regularization}
\label{s:regularisation}

Equation (\ref{e:Phat2}) is a formal description for estimating the
angular structure $\hat{\cal P}_\alpha$ of a gravitational-wave
background.  We refer to this as deconvolution since it tries to
remove the smoothing introduced by the point spread function.
Deconvolution requires inverting the Fisher matrix
$\Gamma_{\alpha\beta}$.  However, in practice, the Fisher matrix
$\Gamma_{\alpha\beta}$ is somewhat ill-conditioned. There are two
reasons for this.

First, the detector pair is diffraction limited. Thus, as we choose a
basis with higher spatial resolution, the condition number of the
Fisher matrix $\Gamma_{\alpha\beta}$ gets worse, resulting in a
reduced signal-to-noise ratio for the deconvolved map.  We can address this
by picking a basis with a reasonable resolution cut-off, which makes
the spherical harmonics basis set with $l \leq l_{\rm max}$ a
natural candidate.

Second, there are certain power distributions $\hat{\cal P}_\alpha$ to
which the detector pair is essentially blind.  For those distributions
positive and negative contributions from different sky locations to
the total cross-correlation essentially cancel. Mathematically they
are described by
\begin{equation}
\label{eq:problematic}
X_\alpha = \Gamma_{\alpha\beta} \hat{\cal P}_\beta \approx 0,
\end{equation}
i.e., they are the eigenfunctions of the Fisher matrix
$\Gamma_{\alpha\beta}$ with the smallest eigenvalues. These
eigenfunctions tend to have $z$-axis rotational symmetry because the
detector pair is rotating with the Earth.  However this symmetry can
be broken by daily variations in detector sensitivity.
To address this second issue, we have chosen to use a singular value
decomposition (SVD) regularization scheme, which we describe in
some detail below.

\begin{figure}
  \psfig{file=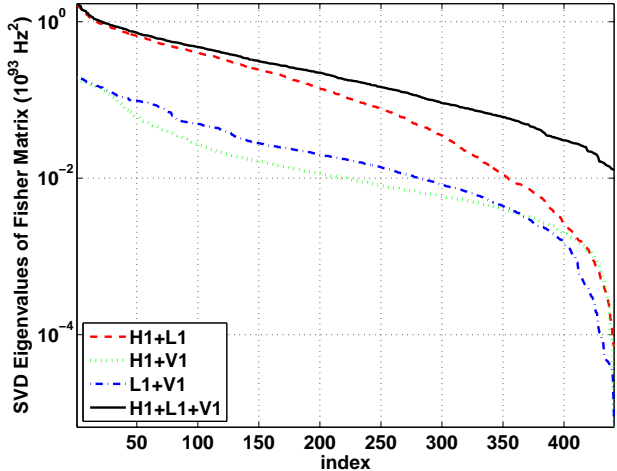,width=3.25in}
  \caption{Eigenvalues of typical Fisher matrices
  $\Gamma_{\alpha\beta}$ for different baselines and the 
  multibaseline detector network.
  For this analysis $l_{\rm max}=20$, corresponding 
  $(l_{\rm max}+1)^2=441$ total modes.
  For each individual baseline some of the SVD eigenmodes
  are (almost) null [see Sec.~\ref{s:regularisation}]. The
  multibaseline network,
  however, has fewer null modes, illustrating the fact that a network
  of detectors acts as a natural regularizer -- independent baselines
  tend to complement each other. The plot was produced using the
  simulated data described in Sec.~\ref{s:simulations}.
  \label{fig:GammaEigenval}}
\end{figure}

Since $\Gamma_{\alpha\beta}$ is Hermitian, its SVD has the form
\begin{equation}
\Gamma = U S U^*\,,
\end{equation}
where $U$ is a unitary matrix and $S={\rm diag}(s_i)$ is a diagonal
matrix with non-negative entries $s_i$ (the eigenvalues of
$\Gamma$). Without loss of generality we can further assume that the
diagonal elements of $S$ are sorted from the largest to 
smallest values.  Then the
problematic modes according to Eq.~(\ref{eq:problematic})
correspond to the last entries of the diagonal of $S$.  Figure \ref{fig:GammaEigenval}
shows the relative size of the eigenvalues of a typical
$\Gamma_{\alpha\beta}$ matrix in the spherical harmonics basis
(with $l_{\rm max}=20$), taken from the no-injection simulation of
Section \ref{s:simulations}.  We can now set
a threshold $s_{\rm min}$ on the size of the eigenvalues, setting all
eigenvalues $s_i < s_{\rm min}$ to infinity (their inverse to zero),
or alternatively to $s_i = s_{\rm min}$. Using this modified matrix
$S'$ we can then define the regularized $\Gamma'$ as
\begin{equation}
\Gamma' = U S' U^*
\end{equation}
and its inverse as
\begin{equation}
\Gamma'^{-1} = U S'^{-1} U^*\,.
\end{equation}
The threshold $s_{\rm min}$ is chosen by weighting the quality
of the deconvolution (larger point spread function for higher values
of $s_{\rm min}$) against the addition of noise due to poorly measured
modes (lower values of $s_{\rm min}$). 
While one can make this trade-off argument
more quantitative, the choice will be somewhat influenced by the spatial shapes
one is looking for. For the purpose of this paper, we simply chose
to keep $2/3$ of all eigenmodes, and set all the small 
eigenvalues equal to $s_{\rm min}$.
 This is somewhat arbitrary, but a reasonable choice
to get rid of the extremely small eigenvalues of a typical Fisher matrix (Figure \ref{fig:GammaEigenval}).
As can be seen in Sec.~\ref{s:simulations}, 
this choice allows for a reasonable recovery of
simulated injections.

\begin{figure}
  \psfig{file=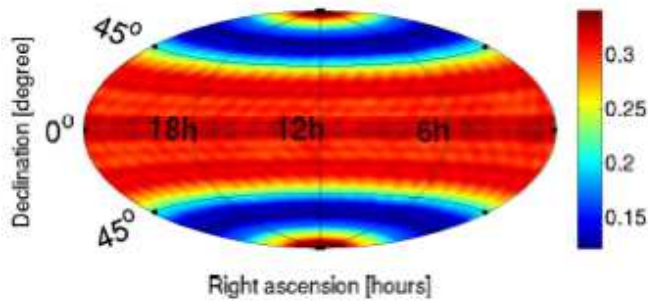,width=3.5in}
  \caption{ Magnitude of the bias due to the SVD regularization
  scheme, for the case of point sources.  A value of 0 implies that
  the expectation value of the corresponding pixel is equal to the
  point source signal strength, while 1 implies that a point source at
  that location would not be seen.
  \label{fig:pixbias}}
\end{figure}

Using this regularization scheme has two side effects that need to be
mentioned.  First, Eq.~(\ref{e:Phat}) is replaced by
\begin{equation}
\hat{\cal P'}_\alpha = (\Gamma'^{-1})_{\alpha\beta}\, X_\beta.
\label{e:Phatreg}
\end{equation}
Thus the expectation value of $\hat{\cal P'}_\alpha$ is
\begin{equation}
\langle \hat{\cal P'}_\alpha\rangle = (\Gamma'^{-1})_{\alpha\beta}
\Gamma_{\beta\gamma} {\cal P}_\gamma \neq {\cal P}_\alpha \,.
\end{equation}
This constitutes a bias in the estimator, which is expected since we
chose to ignore the modes of ${\cal P}_\alpha$ that are poorly
measured.  Under the assumption that we know the shape of the source
this bias can be calculated. Assuming the signal consists of point
sources, Figure \ref{fig:pixbias} shows the size of that bias as a
function of sky position.  
Second, the covariance matrix of
$\hat{\cal P'}_\alpha$ (in the weak-signal approximation) is now given
by
\begin{equation}
\langle \hat{\cal P'}_\alpha \hat{\cal P'}^*_\beta\rangle -
\langle \hat{\cal P'}_\alpha\rangle
\langle \hat{\cal P'}^*_\beta\rangle
= 
(\Gamma'^{-1})_{\alpha\gamma} \Gamma_{\gamma\delta} 
(\Gamma'^{-1})_{\delta\beta}
\,.
\label{e:<PP>reg}
\end{equation}

Finally, we note that adding additional detector pairs
with different baselines can, to a certain degree, 
act as a natural
regulator, simply because one detector network might be more
sensitive to a particular mode than another as illustrated in 
Figure~\ref{fig:GammaEigenval}.
This is described in more detail in the following subsection.

\subsection{Multiple baselines}
\label{s:multiple_baselines}


As shown explicitly in \cite{mitra-et-al} for
the case of the directed radiometer method, the above
analysis can
easily be extended to a network of three or more
detectors with uncorrelated detector noise.
One simply adds the dirty maps $X_\alpha^{IJ}$
and Fisher matrices $\Gamma_{\alpha\beta}^{IJ}$
for each distinct detector pair $IJ$:
\begin{eqnarray}
X_\alpha^{\cal N} =
\sum_I\sum_{J>I}
X_\alpha^{IJ}
\,,
\quad
\Gamma_{\alpha\beta}^{\cal N} =
\sum_I\sum_{J>I}
\Gamma_{\alpha\beta}^{IJ}
\,,
\end{eqnarray}
where the subscript ${\cal N}$ signifies a {\it network} of baselines.
This follows from extending the likelihood formulation in 
Sec.~\ref{s:MLestimation} to include sums over baselines 
as well as frequency and time.
The maximum-likelihood estimators $\hat{\cal P}_\alpha$ then
retain the same form as for the single baseline case, namely
\begin{equation}
\hat{\cal P}_\alpha \ = \ [(\Gamma^{\cal N})^{-1}]_{\alpha\beta} \, X_\beta^{\cal N} \,.
\end{equation}
This follows immediately from Eq.~\ref{e:FisherPartial}.



Different baselines in the network partly complement each other and help fill gaps in sensitivity present in individual baselines pairs.
This has an important consequence.
The sensitivity gaps correspond to degeneracies in the Fisher information matrix, which make it hard to estimate the true stochastic background.
By filling these gaps the network acts as a natural regularizer, as illustrated in Fig.~\ref{fig:GammaEigenval}.

\begin{figure}
  \psfig{file=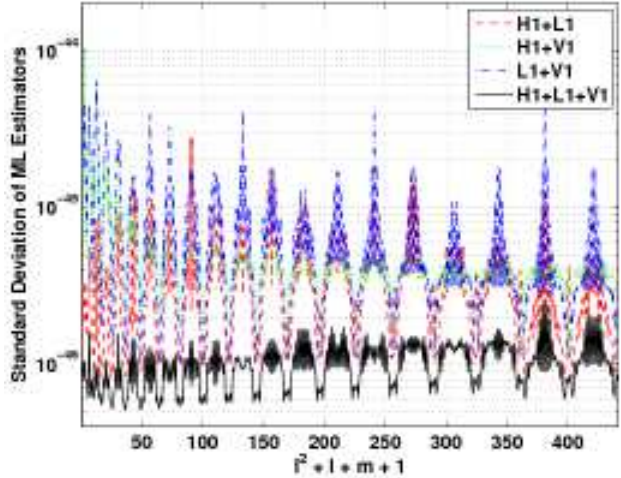,width=3.25in}
  \caption{Standard deviation for spherical harmonics components, without SVD regularization. It illustrates how the multiple baselines (solid line) reduces the estimation error by natural regularization. \label{fig:SD}}
\end{figure}

\begin{figure}
  \psfig{file=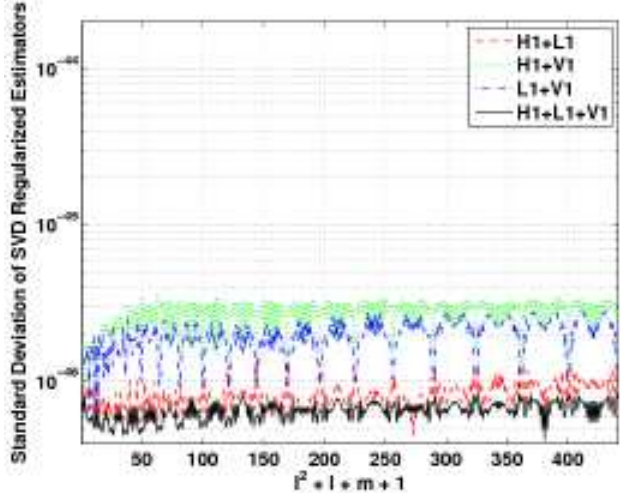,width=3.25in}
  \caption{Standard deviation for spherical harmonics components, with the SVD regularization described in Sec.~\ref{s:regularisation}. While the combination of network and SVD regularization leads to the minimum estimation error, the SVD regularization alone can significantly reduce the estimation error for a single baseline. \label{fig:REGSD}}
\end{figure}


Our next step is to quantify how the natural network regularization and the SVD regularization reduce the error in estimating the spherical harmonic 
components. The full covariance matrix 
of the estimated multipoles $\hat{{\cal P}}_{lm}$ is the best measure of estimation error, but it is inconvenient to compare covariance matrices. Rather, we use the standard deviation of each multipole 
as our figure of merit for estimation error:
\begin{equation}
\sigma_{lm} = \sqrt{{\rm Var}(\hat{\cal P}_{lm})} = \sqrt{[\Gamma^{-1}]_{lm,lm}}
\end{equation}
for unregularized estimators, 
and
\begin{equation}
\sigma'_{lm} = \sqrt{{\rm Var}(\hat{\cal P'}_{lm})}= \sqrt{[(\Gamma')^{-1} \Gamma (\Gamma')^{-1}]_{lm,lm}}
\end{equation}
for regularized estimators
(no summation over $lm$ in either of these two formulas).
We plot the standard deviations for each multipole for both unregularized and regularized estimators
in Fig.~\ref{fig:SD} and Fig.~\ref{fig:REGSD}, respectively. 
Both of these figures indicate that multiple baselines vastly reduce the estimation error. In addition, Fig.~\ref{fig:REGSD} illustrates that though the estimation error is minimized when both the network and SVD regularization are present, the SVD regularization alone can significantly reduce the estimation error (to just
$\sim 25\%$
more than the regularized network error) for the LIGO only baseline.
Thus, even if one interferometer is not unusable for some period, the regularized spherical harmonic moment estimators for the remaining baseline can still provide reasonable results.

\subsection{Model selection for spherical harmonic decomposition}
\label{s:modelselection}

In addition to choosing the cutoff for the SVD regularization
of the Fisher matrix (as described in Sec.~\ref{s:regularisation}),
one needs to specify the value of $l_{\rm max}$, 
the maximum value of the spherical
harmonic index $l$ used in the spherical harmonic decomposition.
Choosing $l_{\rm max}$ fixes the total number of multipole 
moments ${\cal P}_{lm}$,
and hence defines the signal model.
Larger values of $l_{\rm max}$ mean finer angular resolution of
the sky maps and more parameters available to fit the data.
But since the estimators 
$\hat{\cal P}_{lm}$ are correlated with one another,
increasing the number of parameters simultaneously increases 
the uncertainty associated with each parameter.
Thus, there is a tradeoff between accurately modeling the data 
(more parameters) and minimizing uncertainties (fewer parameters).
In this subsection we outline how Bayesian model selection can be used to fix $l_\text{max}$.
(This discussion is meant to motivate future study as we do not implement a model selection scheme in this work.)

Bayesian model selection (see, e.g., \cite{Sivia:1996}) is a
framework in which the data themselves determine which 
signal model is most appropriate.
The basic idea is to compare the various models
(e.g., $M_1$ and $M_2$) by computing the ratio of the 
probability of the models given the data $D$.
By Bayes' theorem, we have
\begin{equation}
\frac{p(M_1|D)}{p(M_2|D)} = 
\frac{p(D|M_1)}{p(D|M_2)}\frac{p(M_1)}{p(M_2)}
\end{equation}
where $p(M_1)$ and $p(M_2)$ are the a~priori 
probabilities of the models, and 
$p(D|M_1)$ and $p(D|M_2)$ are the likelihood functions
for the data given the two models.
The ratio of the likelihoods $p(D|M_1)/p(D|M_2)$ is known as 
the ``Bayes factor'' (see, e.g.,~\cite{Sivia:1996}).
If there is no a~priori reason to prefer one model over
the other (as is often the case), then $p(M_1)/p(M_2)=1$,
implying that the posterior odds 
is just the ratio of the likelihood functions,
$p(D|M_1)/p(D|M_2)$. 
Since a given model often involves a set of parameters
$a$, calculating the likelihood of the data for a
given model requires {\em marginalizing} over the possible
values of these parameters---i.e.,
\begin{equation}
p(D|M) = \int da\> p(D|a,M)p(a|M)
\,,
\end{equation}
where $p(a|M)$ is the prior probability distribution of 
the parameters for that model.

In situations where the data is informative---i.e., when the 
likelihood function $p(D|a,M)$  
is peaked relative to the prior $p(a|M)$---we have 
the approximate relation
\begin{equation}
p(D|M)\approx 
p(D|\hat a,M)\frac{\delta a}{\Delta a} 
\,,
\label{e:occam_approx}
\end{equation}
where $\hat a$ is the value of $a$ that maximizes the 
likelihood,
$\delta a$ is the range 
of parameter values over which the likelihood is peaked, and 
$\Delta a$ is the full range of parameter values.
The factor $\delta a/\Delta a$ penalizes a model that uses
more parameter space volume than needed to fit the data.
This factor can be understood in terms of {\em Occam's razor}, 
which says that 
everything else being equal, simpler models that can 
adequately fit the data are preferred.

For example, if we ignore the subtleties described in
Sec.~\ref{s:regularisation} related to the
inversion of the Fisher matrix $\Gamma_{lm,l'm'}$,
an $l_\text{max}=30$ map (961 parameters) will always
fit the data better than an 
$l_\text{max}=5$ map (36 parameters) in the sense
of having a larger value of $p(D|\hat a,M)$.
One can imagine, however, a stochastic background 
characterized by ${\cal P}_{lm}$ up to only $l_\text{max}=5$.
In that case, an $l_\text{max}=30$ fit would introduce a great 
many unnecessary parameters, which means a much
smaller value of $\delta a/\Delta a$ 
offsetting the larger value of $p(D|\hat a,M)$.
In addition, since the multipole moment estimators for
different $l$ and $m$ are correlated with one another, choosing
a large value of $l_{\rm max}$ would have the undesirable 
effect of worsening the uncertainty associated with the 
$\hat {\cal P}_{lm}$ up to $l_\text{max}=5$.

In the context of our search for an anisotropic stochastic
gravitational-wave background, the model $M$ is that a 
signal is present having  multipole moments up to $l_{\rm max}$.
The data are the measured cross-spectra $C_{ft}$, and
the likelihood function is given by 
Eq.~(\ref{e:likelihood}), where the signal model is
defined by multipole moments up to $l_{\rm max}$.
Thus, the quantity we need to calculate is the marginalized
likelihood
\begin{align}
&p(\{C_{ft}\}|l_{\rm max}) =
\nonumber
\\
&\quad
\int d\{{\cal P}_{lm}\}\,
p(\{C_{ft}\}|\{{\cal P}_{lm}\}, l_{\rm max})
p(\{{\cal P}_{lm}\}|l_{\rm max})
\,,
\end{align}
where $p(\{{\cal P}_{lm}\}|l_{\rm max})$ are the prior
probability distributions for the multipole moments.
Since the parameter space is large 
(a total of $(l_{\rm max}+1)^2$ parameters), 
sophisticated Markov Chain Monte Carlo techniques (see \cite{Skilling:2006})
may be required to numerically evaluate this integral.
If the data turn out to be informative, then one has the
much simpler expression
\begin{equation}
p(\{C_{ft}\}|l_{\rm max}) 
\approx
p(\{C_{ft}\}|\{\hat{\cal P}_{lm}\}, l_{\rm max}) 
\frac{\sqrt{\det(\Gamma^{-1})}}
{\prod_{lm}\Delta{\cal P}_{lm}}
\label{e:pbayes}
\end{equation}
where $\hat{\cal P}_{lm}$ are the maximum-likelihood estimators
given by Eq.~(\ref{e:Phat}),
and ${\Delta {\cal P}_{lm}}$ are characteristic widths 
of the prior distributions.
Whether or not one can use this approximation depends on the
actual data and the choice of priors.
In practice, it may be possible to use limits from 
previous, less sensitive analyses to set the widths 
of the priors for (at least some of) the ${\cal P}_{lm}$.
In the absence of strong a~priori knowledge, the widths 
of the priors will necessarily be large, reflecting our
uncertainty in the values of the signal parameters.

\section{RESULTS OF SIMULATIONS}
\label{s:simulations}

In this section, we present the results 
produced by our data analysis code 
for simulated stochastic signals injected into 
simulated detector noise.
We focus attention on analyses done in the spherical
harmonics basis, as similar studies for the pixel-based
decomposition have already beeen discussed in detail 
in the context of the radiometer
analysis~\cite{mitra-et-al,radiometer}. 
We find that the spherical harmonic analysis method can 
successfully recover simulated signals injected 
into simulated noise for several different types of 
stochastic gravitational-wave
backgrounds, e.g., isotropic sources, 
dipole sources, point sources, diffuse sources, etc..
We also verify that the results of the standard 
isotropic and radiometer
analyses are recovered 
as special limiting cases of the spherical 
harmonic decomposition analysis for
$l_\text{max}=0$ and $l_\text{max}\rightarrow \infty$, 
respectively.

\subsection{Simulation details}

The simulations described in this section are made
up of twenty-four jobs, each consisting of 
approximately one hour of data.
Since the beam pattern matrix of the detector 
varies with local sidereal time, we chose 
the start time of each job so that the data are distributed 
(nearly) uniformly over a sidereal day.
The twenty-four jobs are further broken down into 
one-minute segments (so $\tau=\unit[60]{sec}$), on which 
the analysis described 
in Section~\ref{s:MLestimation} is then applied.

\begin{figure}
  \psfig{file=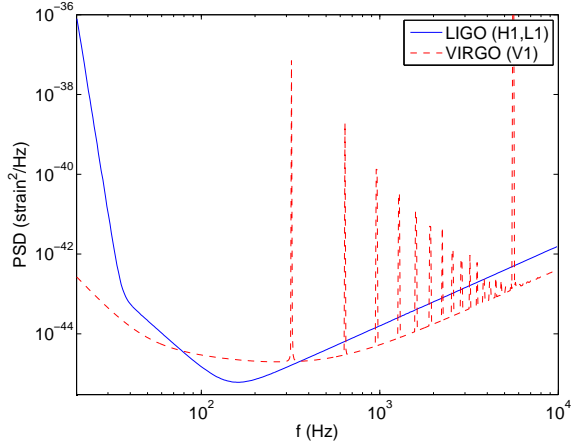,width=3in} 
  \caption{The design power spectral densities used
  to simulate detector noise for the 
  LIGO $\unit[4]{km}$ interferometers (H1 and L1) 
  and the $\unit[3]{km}$ Virgo interferometer (V1)~\cite{S5detector,Virgo-PSD}.
  \label{fig:PSDs}}
\end{figure}

The simulated time-series data 
are sums of simulated detector 
noise and simulated stochastic signals for
several different angular distributions.
The simulated detector noise are constructed so as 
to reproduce (on average) the design power spectral 
densities of the different detectors---in our case,
the $\unit[4]{km}$ Hanford and Livingston LIGO interferometers
(H1 and L1) and the 3-km Virgo interferometer (V1). 
See Figure~\ref{fig:PSDs}. 
The simulated stochastic signals we consider include:
no injection (i.e., just detector noise), 
an isotropic (i.e., monopole) source,
a dipole source, a point source, two point sources, a 
diffuse source clustered around the galactic plane, and 
a diffuse source clustered around $\text{dec}=0^\circ$.
Note that the dipole source is injected on top of a monopole of 
twice its amplitude, so that the signal power is positive 
everywhere on the sky.

The spectral shape of the stochastic signal is taken to 
be constant ($\bar H(f)=1$) for all the injections and 
for all the analyses.
The overall amplitude of the signals are different for 
the different injections, chosen to be large enough to be easily detectable
in one sidereal day of total integration time.\footnote{Although the auto-correlated power spectra for some of the injected signals were {\em larger} than those for the simulated detector noise, we could still use the weak-signal formulas from Section~\ref{s:MLestimation} since the {\em cross-correlated} gravitational-wave power was always much less than the auto-correlated power in the detector output (which consists of detector noise {\em plus} the auto-correlated signal). The reduction in the cross-correlated signal power is due to the overlap factors $\gamma_\alpha(f,t)$ for physically separated detectors being much smaller in magnitude than those for the same detector.}
Table~\ref{tab:signalstrengths} 
lists the expected values of 
${\cal P}_{00}/\sqrt{4\pi}$ and 
${\cal P}(\hat\Omega)\big|_{\rm max}$,
either of which fix the scale of the various injections.
The factor of $1/\sqrt{4\pi}$ multiplying ${\cal P}_{00}$ 
is included to allow direct comparison with the sky map 
plots of 
${\cal P}(\hat\Omega)=\sum {\cal P}_{lm}Y_{lm}(\hat\Omega)$
shown later in this section, noting that 
$Y_{00}(\hat\Omega)=1/\sqrt{4\pi}$.
The maximum power values are given for easy comparison for the point source injections.

\begin{table}
  \begin{tabular}{l|c|c}
    \hline\hline
    Injection type & 
    ${\cal P}_{00}/\sqrt{4\pi}$ &
    ${\cal P}(\hat\Omega)\big|_{\rm max}$ \\
    & (strain${}^2$/Hz/rad${}^2$) & (strain${}^2$/Hz/rad${}^2$) \\ 
    \hline
    Monopole & $5.6\times 10^{-45}$ & $5.6\times 10^{-45}$ \\
    Dipole & $1.1\times 10^{-44}$ & $2.1\times 10^{-44}$ \\
    1 point source & $1.6\times 10^{-47}$ & $4.1\times 10^{-45}$ \\
    2 point sources & $3.2\times 10^{-47}$ & $4.0\times 10^{-45}$ \\
    Diffuse source (galactic) & $3.8\times 10^{-45}$ & $2.0\times 10^{-44}$ \\
    Diffuse source (${\rm dec}=0^\circ$) & $4.2\times 10^{-45}$ & $2.0\times 10^{-44}$ \\
    \hline\hline
    \end{tabular}
  \caption{Expected values of ${\cal P}_{00}/\sqrt{4\pi}$ and 
  ${\cal P}(\hat\Omega)\big|_{\rm max}$ for the different injections.
           \label{tab:signalstrengths}}
\end{table}

The analysis code was then run on the 
simulated data, 
decomposing the relevant quantities with respect
to the spherical harmonic basis as described in 
Section~\ref{s:MLestimation}.
The main output of the analysis for a 
particular simulation are the spherical harmonic
components of the dirty map $X_{lm}$ and the 
beam matrix $\Gamma_{lm,l'm'}$.
The maximum-likelihood estimates of the true 
multipole moments ${\cal P}_{lm}$ of the 
gravitational-wave sky
are then obtained by inverting the beam matrix
$\Gamma_{lm,l'm'}$ 
(either with or without SVD as discussed in
Sec.~\ref{s:regularisation}), and then applying
that inverse to the $X_{lm}$ to get the 
components of the clean map $\hat{\cal P}_{lm}$.
(In what follows, a `clean map' will mean the
map constructed from the regularized inverse,
unless we explicitly indicate otherwise.)

\subsection{Comparison with previous searches}

As an initial check of the analysis pipeline, 
we verified that the spherical harmonic decomposition
code reproduced the results of the standard 
isotropic~\cite{S1Isotropic,S3Isotropic,S4Isotropic} and radiometer~\cite{mitra-et-al,radiometer} analyses in the limits
$l_\text{max}=0$ and $l_\text{max}\rightarrow\infty$,
respectively.
For the isotropic comparison,
we analyzed simulated data with no injection 
(just detector noise) and compared the 
$l_\text{max}=0$ results 
(i.e., the maximum-likelihood estimate 
$\hat{\cal P}_{00}$ of the monopole moment, 
and the associated 1-sigma error bar $\sigma_{00}$) 
to an identical analysis 
performed with the isotropic search code.
The results, presented in Table~\ref{tab:isomono}, 
show that the two methods give the same answers
(to round-off error)
for the isotropic component of the background.
\begin{table}
  \begin{tabular}{l|c|c}
    \hline\hline
    Method & $\hat{\cal P}_{00}/\sqrt{4\pi}$ & 
    $\sigma_{00}/\sqrt{4\pi}$ \\
    & (strain${}^2$/Hz/rad${}^2$) & (strain${}^2$/Hz/rad${}^2$) \\ 
    \hline
    isotropic & $4.207339\times10^{-49}$ & $3.209030411\times10^{-48}$ \\
$l_\text{max}=0$ & $4.207328\times10^{-49}$ & $3.209030408\times10^{-48}$ \\
    \hline\hline
    \end{tabular}
  \caption{A comparion of the maximum-likelihood
  estimates and error bars for the spherical harmonic 
  decomposition code ($l_\text{max}=0$) and the 
  standard isotropic search. 
  \label{tab:isomono}}
\end{table}

\begin{figure*}[hbtp!]
  \begin{tabular}{cc}
  \psfig{file=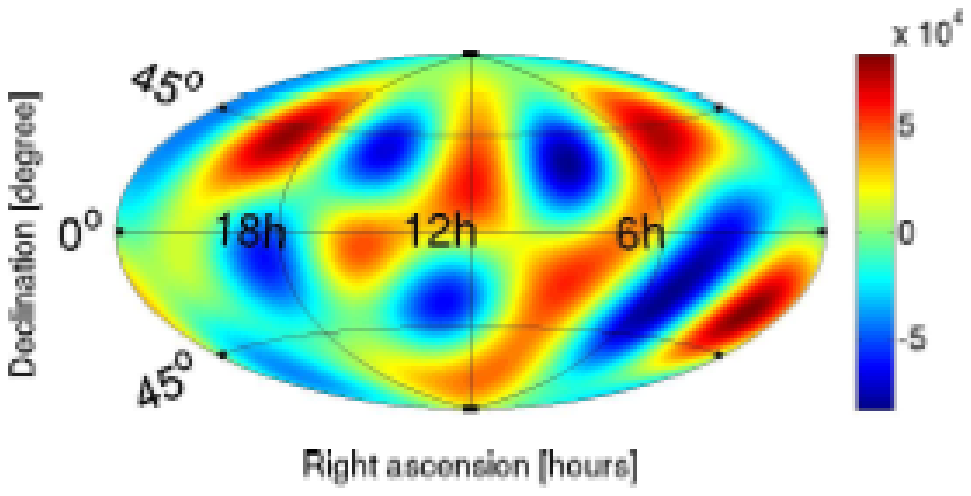,width=3in} &  \psfig{file=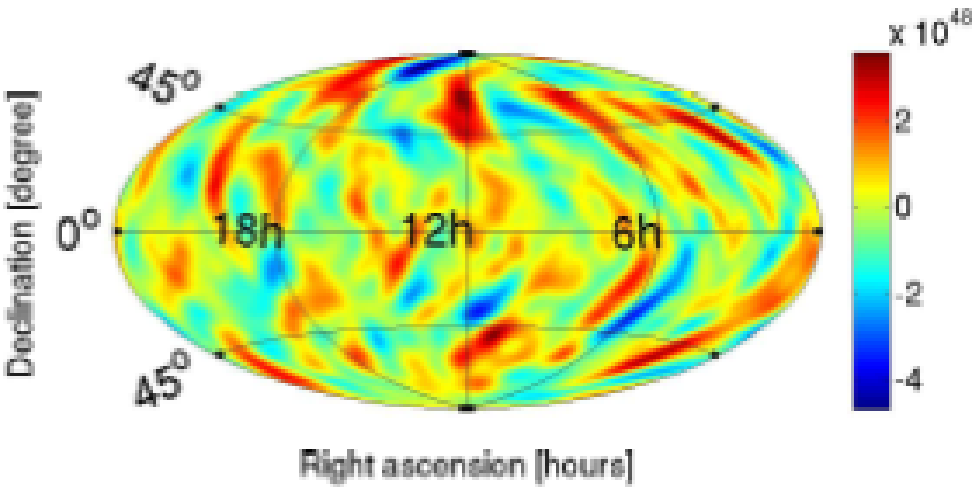,width=3in} \\
  \psfig{file=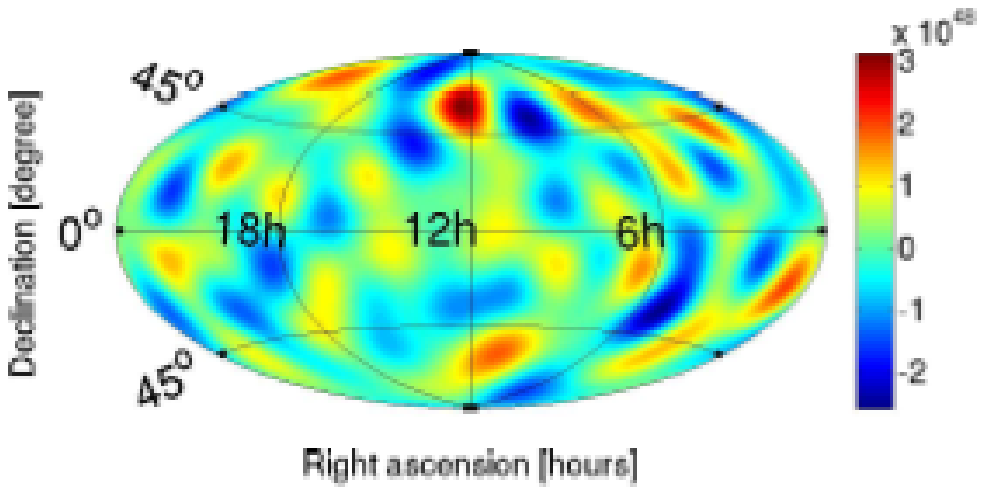,width=3in} & \psfig{file=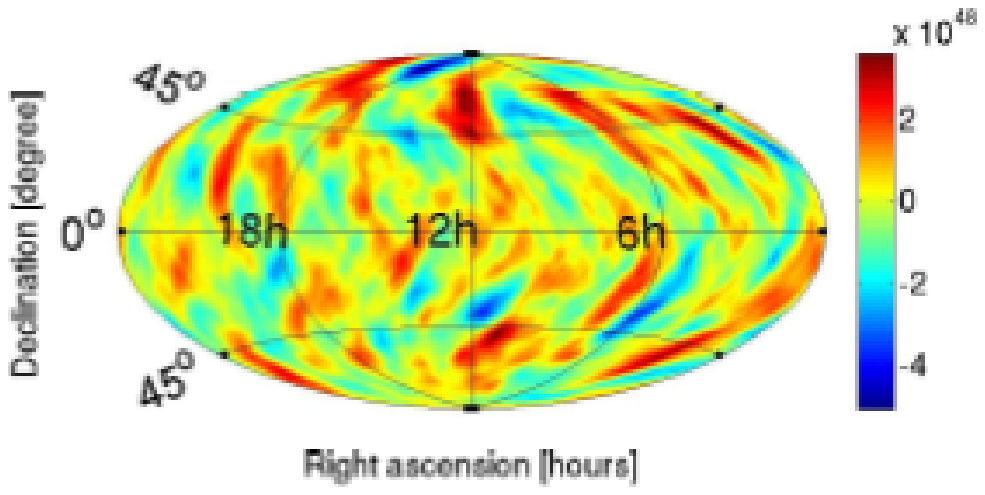,width=3in} \\
  \psfig{file=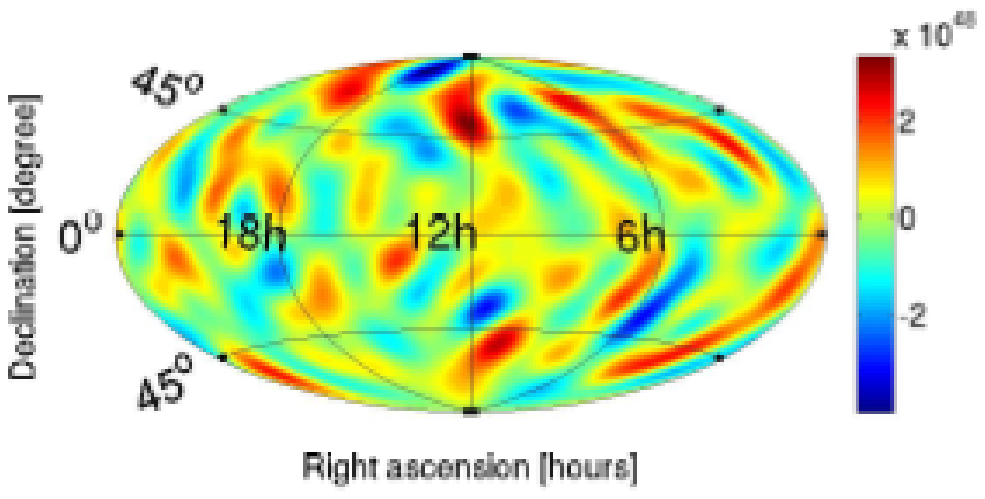,width=3in} & \psfig{file=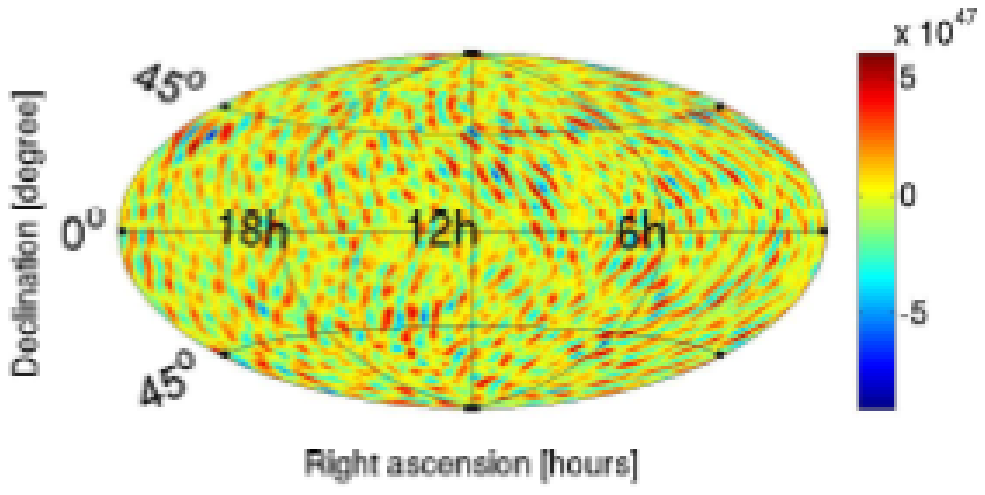,width=3in} \\
  \psfig{file=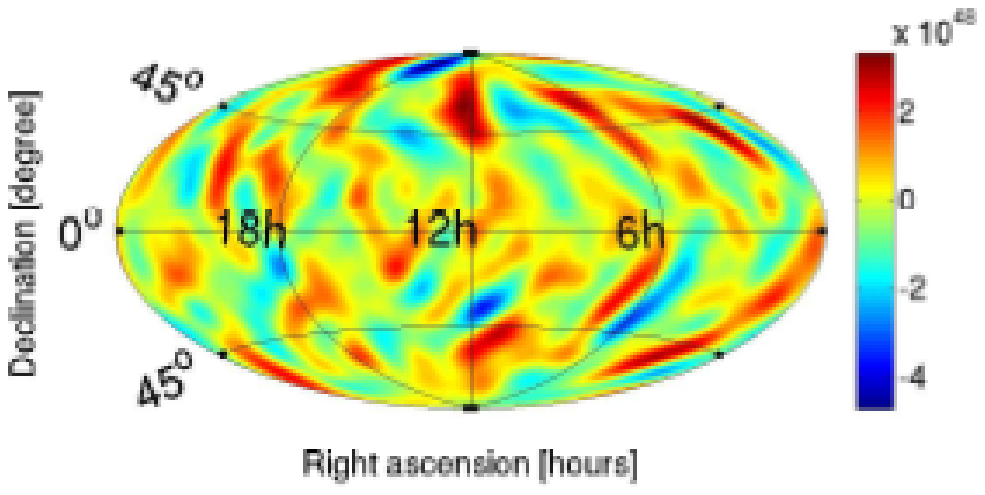,width=3in}  & 
  \end{tabular}
  \caption{Left column from top to bottom: dirty maps of no 
  injection (just detector noise) produced with
  spherical harmonics decomposition code for 
  $l_\text{max}=5$, $l_\text{max}=10$, $l_\text{max}=15$, 
  $l_\text{max}=20$.  Right column: dirty maps with $l_\text{max}=30$, with the radiometer search code, the difference between the $l_\text{max}=30$ map and the radiometer map.  By reading top to bottom, one can see how the spherical harmonic dirty map approaches the radiometer map as $l_\text{max}$ increases.  Residual fluctuations on the difference map appear consistent with the angular scale set by $l_\text{max}=30$.
  \label{fig:radio}}
\end{figure*}

For the radiometer comparison, we analysed the same simulated
data with no injection 
(just detector noise) with both the spherical harmonic
decomposition code for different values of $l_\text{max}$, 
and compared the resultant dirty sky maps
constructed from the $X_{lm}$ with the pixel-based map 
produced by the radiometer search code.  
Figure~\ref{fig:radio} shows that the spherical 
harmonic algorithm successfully reproduces the radiometer 
analysis in the limit of large $l_\text{max}$.
For a radiometer pixelisation appropriate for the 
diffraction limited beam pattern at $f\sim \unit[1]{kHz}$, 
$l_\text{max}=30$ yields a good approximation.
The difference between the $l_{\rm max}=30$ map and 
radiometer map has fluctuations consistent with the 
angular scale set by $l_{\rm max}=30$---i.e., the two 
analyses agree for angular scales accessible up to 
$l_{\rm max}=30$; they differ only for finer angular resolutions.

\begin{figure}[hbtp!]
  \begin{tabular}{c}
  \psfig{file=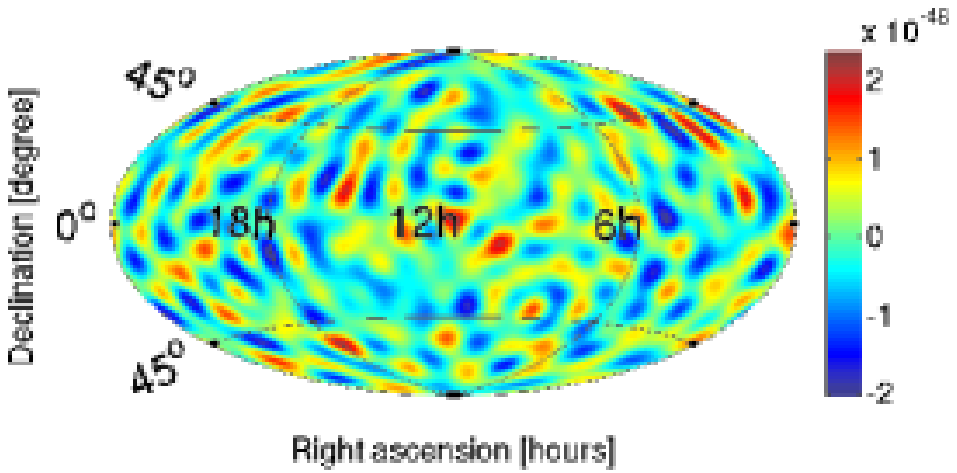,width=3in} \\
  \psfig{file=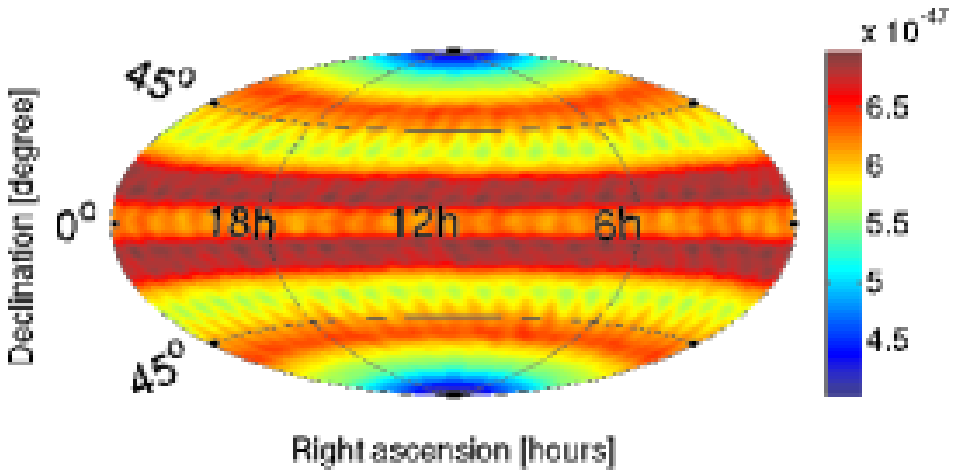,width=3in} \\
  \psfig{file=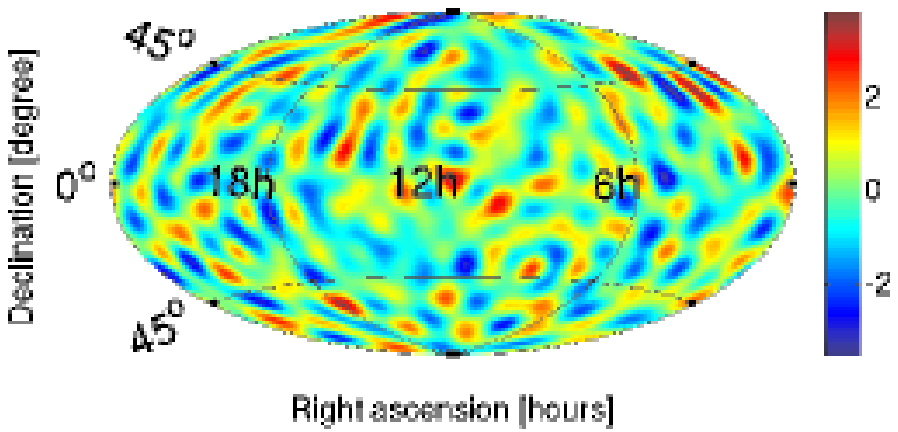,width=3in} \\
  \psfig{file=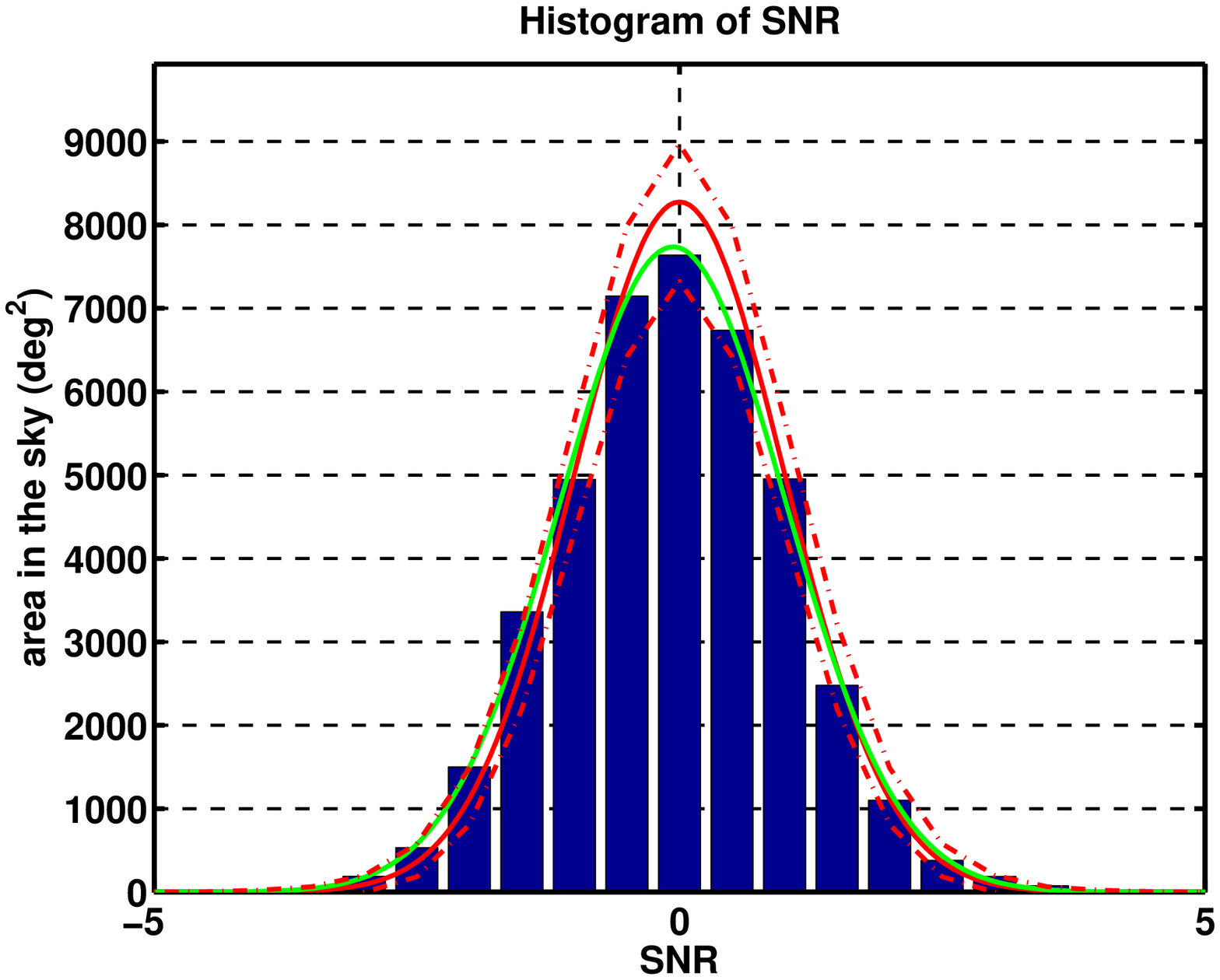,width=3in}
  \end{tabular}
  \caption{From top to bottom: 
  A clean sky map for the case of no injection; 
  next, the associated map of uncertainty, $\sigma_{{\cal P}(\hat{\Omega})}$; 
  next, the associated SNR map; 
  bottom, a histogram of SNR.
  The banded structure on the uncertainty map can be understood in terms of the directional sensitivity of two cross correlated interferometers.
  The rotation of the Earth ensures that the uncertainty is uniform in right ascension for a fixed declination.
  The blue histogram bars are data, the dark red line is a 
  Gaussian fit (sigma=1, mean=1), the light green 
  line is a maximum-likelihood fit (sigma=1.07, mean=0.06), 
  and the dashed line is the 1-sigma error for 400 independent points. 
  $l_\text{max}=20$. \label{fig:SNR}}
\end{figure}

In Figure~\ref{fig:SNR} we show the sky map for this no-injection 
simulation that has been cleaned by the SVD algorithm, 
a SNR map for this clean map, and a histogram of the SNR map.
It is readily apparent that the fluctuations are consistent with detector noise (no signal.)

Thus, the spherical harmonic algorithm reproduces the 
isotropic and radiometer analyses as special cases, 
while allowing us the flexibility to consider more 
general cases, all within a single framework.

\subsection{Sky maps - Single baseline}
\label{s:skymaps-single-baseline}

In this subsection we focus on simulations utilizing the H1-L1 baseline.
The sky maps constructed from the 
$\hat{\cal P}_{lm}$ for the various stochastic
simulations are shown in the following
figures:

1) In the top panel of Figure~\ref{fig:ptsrc} we 
plot a dirty sky map for a point source injection 
with $l_\text{max}=20$.
While the location of the point source at 
$(\text{ra,dec})=(\unit[6]{hr},+45^\circ)$ is readily 
apparent, the source is smeared and it is surrounded 
by artifacts arrising from the beam pattern function.
If we attempt to produce a clean map by naively 
inverting $\Gamma_{\alpha\beta}$ as in the second 
panel of Figure~\ref{fig:ptsrc}, we find that the 
`clean' map is actually worse (less representative 
of the injection) than the dirty map due to 
singularities in the inverted matrix 
(as described in Section~\ref{s:data_analysis}).
In the third panel of Figure~\ref{fig:ptsrc}, we 
present a clean map derived using the SVD algorithm.
The location of the point source is readily apparent, 
and the SVD algorithm has removed some of the artifacts 
and smearing associated with the dirty map.
In the fourth panel we plot the associated SNR map.

When comparing these maps one should bear in mind 
that the clean and dirty maps have different interpretations, 
and so the color scales have very different numerical ranges.
In the illustrative case of $l_\text{max}=0$, for example, 
$X_{00}={\cal P}_{00}/\sigma_{00}^2$, where $X_{00}$ is the dirty map 
and ${\cal P}_{00}$ is the clean map.

\begin{figure}[hbtp!]
  \begin{tabular}{c}
    \psfig{file=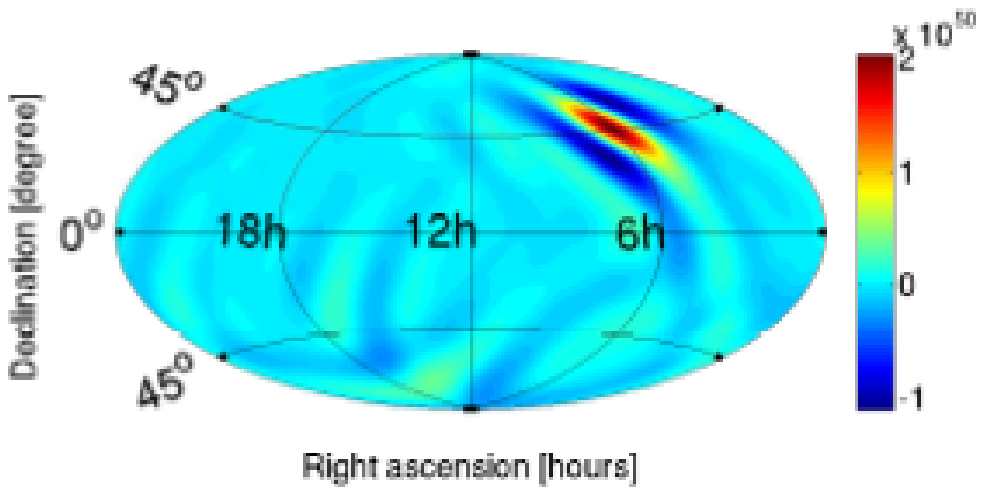,width=3in} \\
    \psfig{file=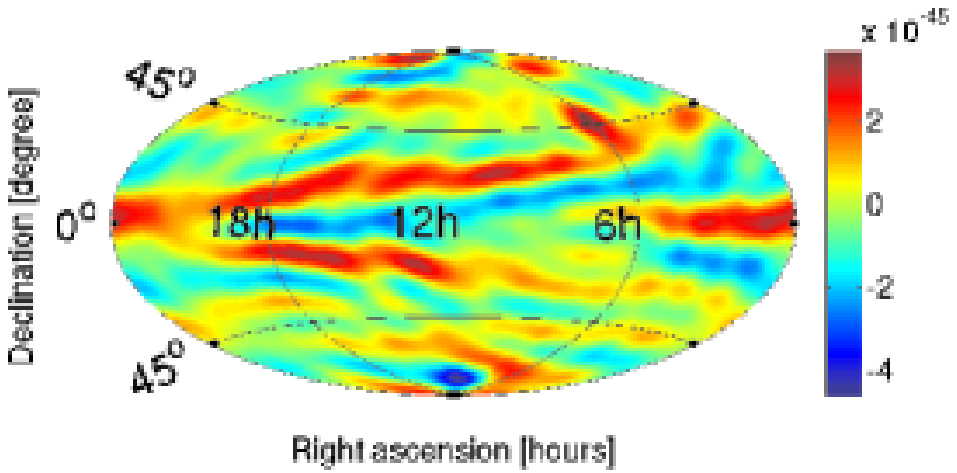,width=3in} \\
    \psfig{file=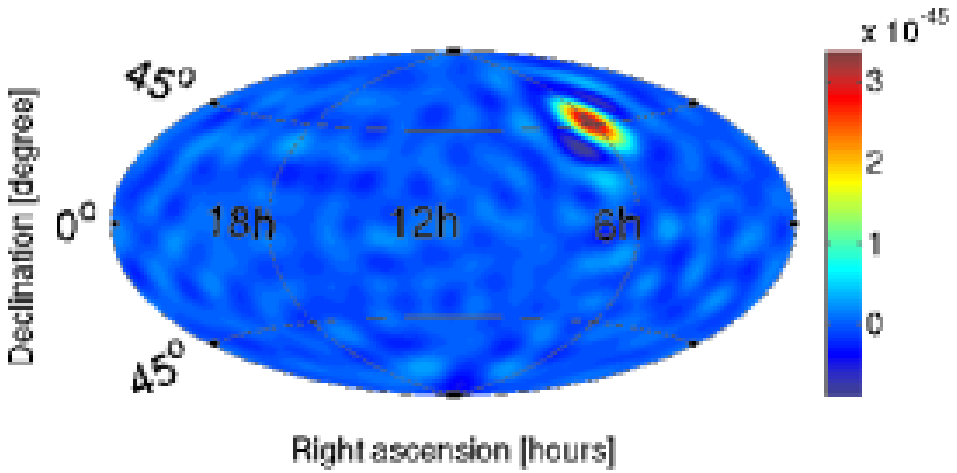,width=3in} \\
    \psfig{file=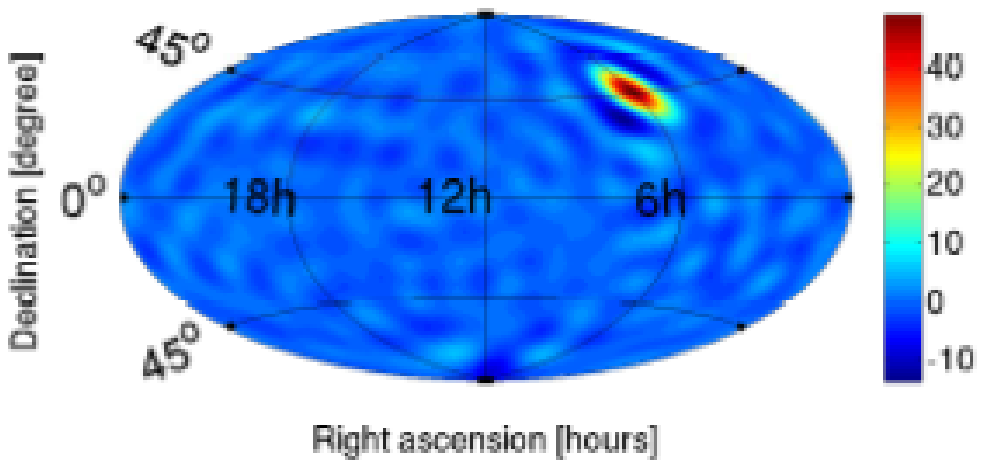,width=3in} 
  \end{tabular}
  \caption{From top to bottom: a dirty sky map of a point source
  injection at $(\text{ra,\,dec})=(\unit[6]{hr},+45^\circ$), a clean
  sky map (without SVD) of the same injection, a clean sky map
  (with SVD) of the same
  injection, a map of SNR. The source has maximum SNR of 49.  $l_\text{max}=20$. \label{fig:ptsrc}}
\end{figure}

\begin{figure}[hbtp!]
  \begin{tabular}{c}
    \psfig{file=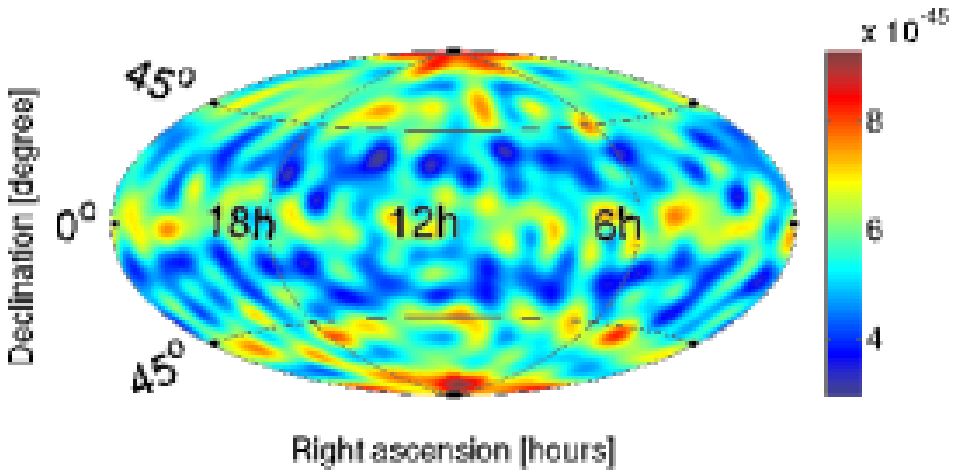,width=3in} \\
    \psfig{file=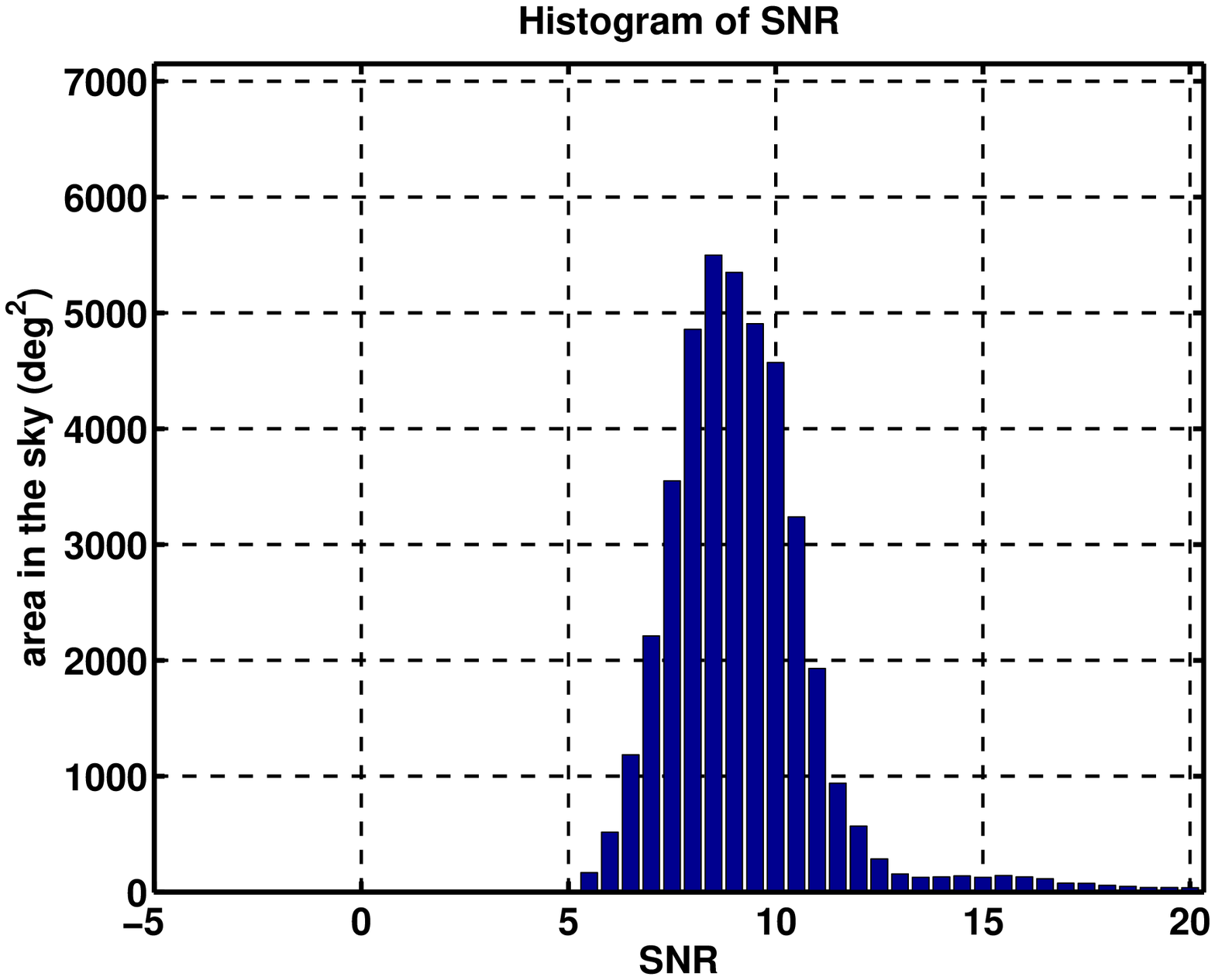,width=3in}
  \end{tabular}
  \caption{Above is a clean sky map of an isotropic injection. Below is a histogram of SNR.  The average SNR across the map is 9.1. $l_\text{max}=20$.
  \label{fig:mono}}
\end{figure}

2) In Figure~\ref{fig:mono} we plot a clean sky map for an isotropic
injection with $l_\text{max}=20$.  For this injection, ${\cal
  P}(\hat{\Omega})=\unit[5.6\times10^{-45}]{strain^2/Hz/rad{}^2}$,
a value which is indicated by green in
Figure~\ref{fig:mono}.

\begin{figure}[hbtp!]
  \psfig{file=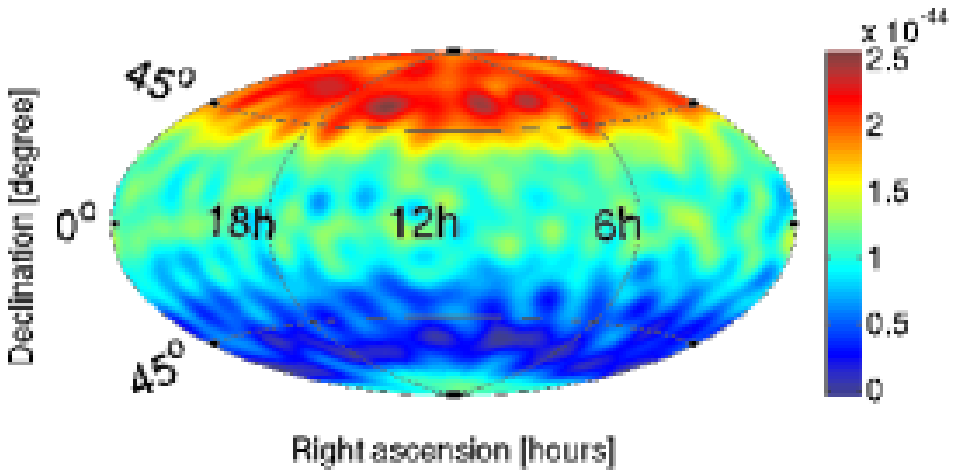,width=3in}
  \caption{A clean map of a dipole injection oriented along the $z$ axis. $l_\text{max}=20$. \label{fig:dipole}}
\end{figure}

3) In Figure~\ref{fig:dipole} we plot a clean sky 
map for a dipole injection with $l_\text{max}=20$.

\begin{figure}[hbtp!]
  \psfig{file=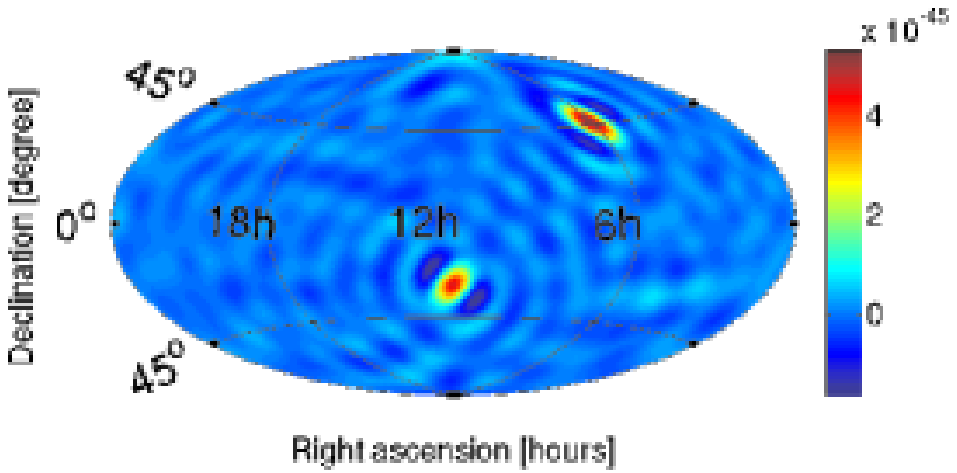,width=3in}
  \caption{A clean sky map of two point sources, one at
  $(\text{ra,\,dec})=(\unit[6]{hr},+45^\circ$) ($\text{SNR}=81$) and the other at
  $(\text{ra,\,dec})=(\unit[12]{hr},-30^\circ)$ ($\text{SNR}=76$).
  $l_\text{max}=20$. \label{fig:twoptsrc_clean}}
\end{figure}

4) In Figure~\ref{fig:twoptsrc_clean} we plot a clean sky map 
for an injection of two point sources with $l_\text{max}=20$.

\begin{figure}[hbtp!]
  \begin{tabular}{c}
    \psfig{file=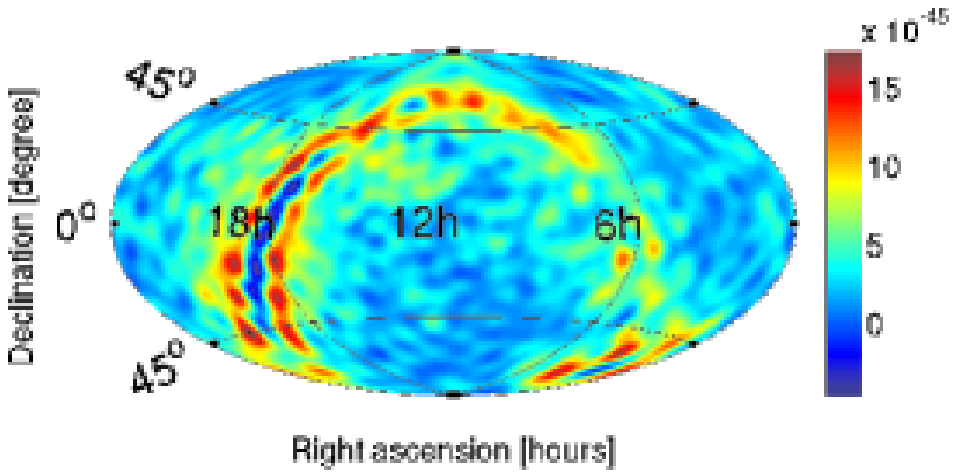,width=3in} \\
    \psfig{file=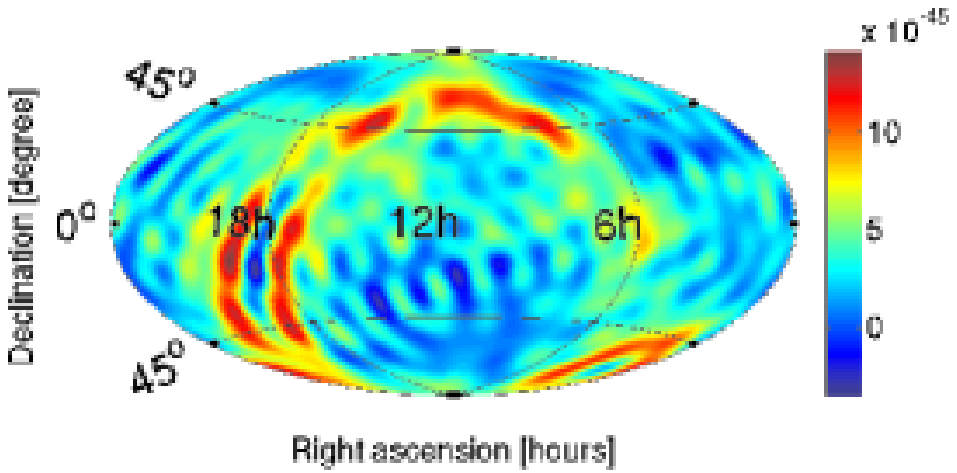,width=3in}
  \end{tabular}
  \caption{Above: a toy model injection corresponding to a map
  measured by the WMAP satellite~\cite{WMAP} meant to mimic a diffuse
  source clustered in the galactic plane ($b=0^\circ$). The map
  utilizes HEALPix~\cite{HEALPix} and the injection was simulated
  using the Planck Simulator~\cite{Planck}. Below: a clean map
  recovered from this injection. $l_\text{max}=20$. \label{fig:eq}}
\end{figure}

5) In the first panel of Figure~\ref{fig:eq} we plot an 
injection of a diffuse source clustered in the galactic plane.
In the second panel we plot the clean sky map recovered 
from this injection using $l_\text{max}=20$.

\begin{figure*}[hbtp!]
  \begin{tabular}{cc}
    \psfig{file=equatorial_injection.eps,width=3in} & 
    \psfig{file=equ_H1L1_clean_20.eps,width=3in}  \\
    \psfig{file=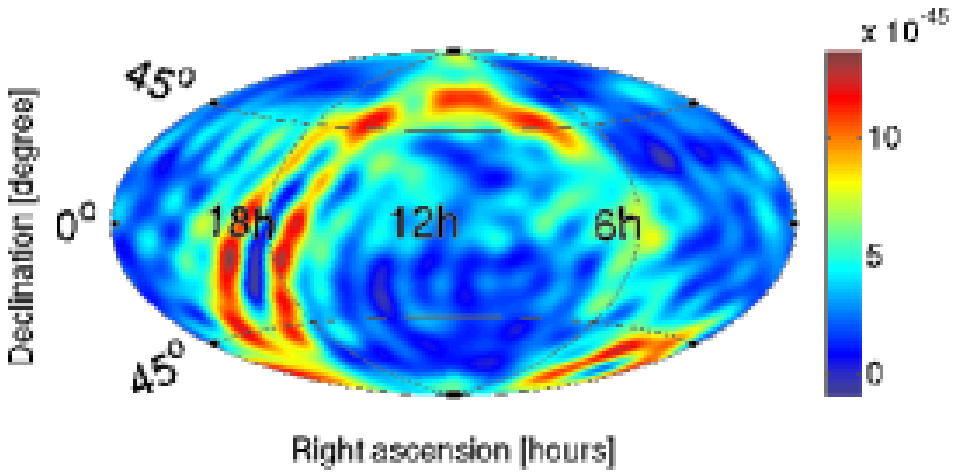,width=3in} & 
    \psfig{file=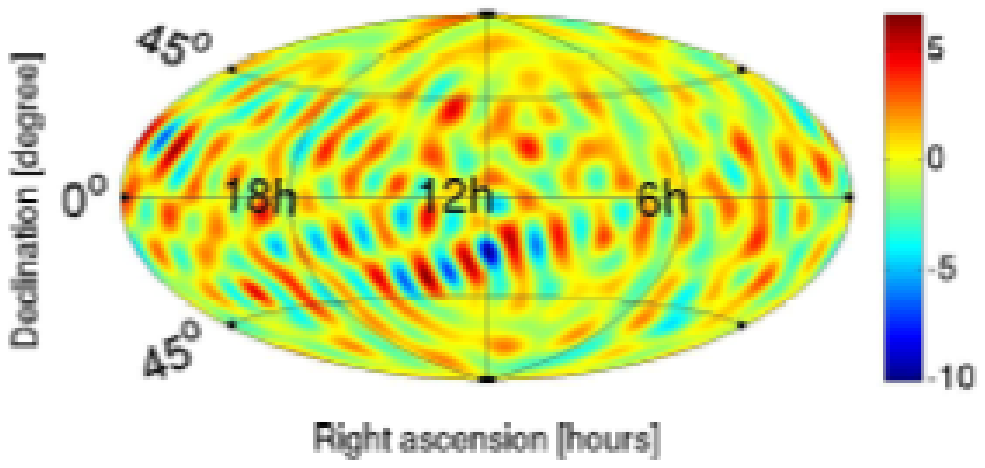,width=3in} \\
    \psfig{file=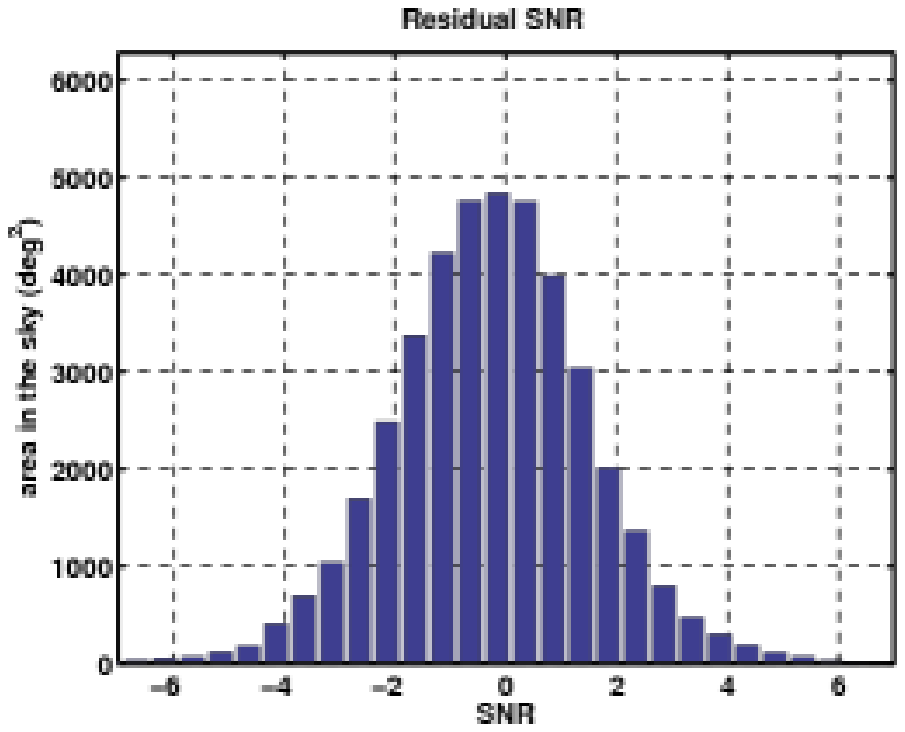,width=3in} &
  \end{tabular}
  \caption{Top-left: the original injection. 
Mid-left: the regularized version of the injection ${\cal
P}^\prime_\alpha$.  
Top-right: the recovered clean map
$\hat{\cal P}^\prime_\alpha$.  
Mid-right: ${\cal P}^\prime_\alpha-\hat{\cal
P}^\prime_\alpha$ normalized by $\sigma_{{\cal P}(\hat{\Omega})}$.  
Bottom-left: a histogram of these residuals.  
The fluctuations in the residuals appear to be consistent with detector noise.  
$l_\text{max}=20$. \label{fig:diff}}
\end{figure*}

6) One way to test that an injection is recovered successfully and
without bias is to plot the injected signal map minus the recovered
clean map.  To do this, we must take into account that the clean map
was produced using SVD (see Section~\ref{s:regularisation}).  That is,
we need to compare the regularized extracted clean map $\hat{\cal
  P}^\prime_\alpha$ with a `regularized' version of the injected map:
\begin{equation}\label{eq:arf}
  {\cal P}_\alpha^\prime = (\Gamma'^{-1})_{\alpha\beta}\,\Gamma_{\beta\,\gamma}\,{\cal P}_\gamma
\,.
\end{equation}
Here $\Gamma_{\beta\,\gamma}$ is the Fisher matrix, 
$(\Gamma'^{-1})_{\alpha\beta}$ is its regularized inverse, and
${\cal P}_\gamma$ is the injected map.
In Figure~\ref{fig:diff} we plot 
${\cal P}^\prime_\alpha-\hat{\cal P}^\prime_\alpha$ 
for the galactic injection depicted in Figure~\ref{fig:eq}.

\begin{figure*}[hbtp!]
  \begin{tabular}{cc}
    \psfig{file=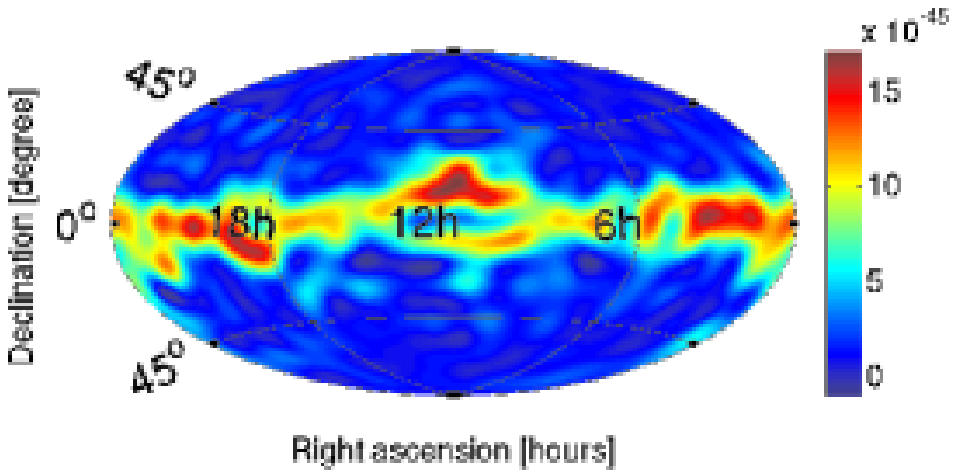,width=3in} & \psfig{file=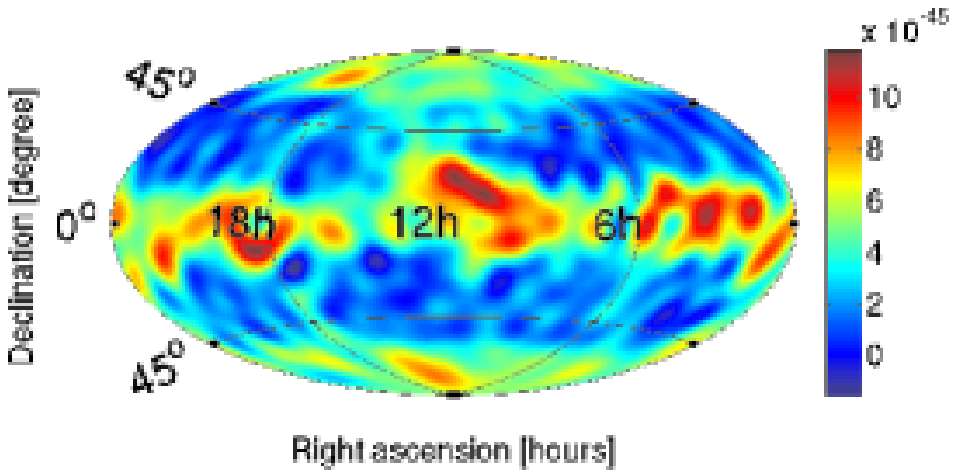,width=3in} \\
    \psfig{file=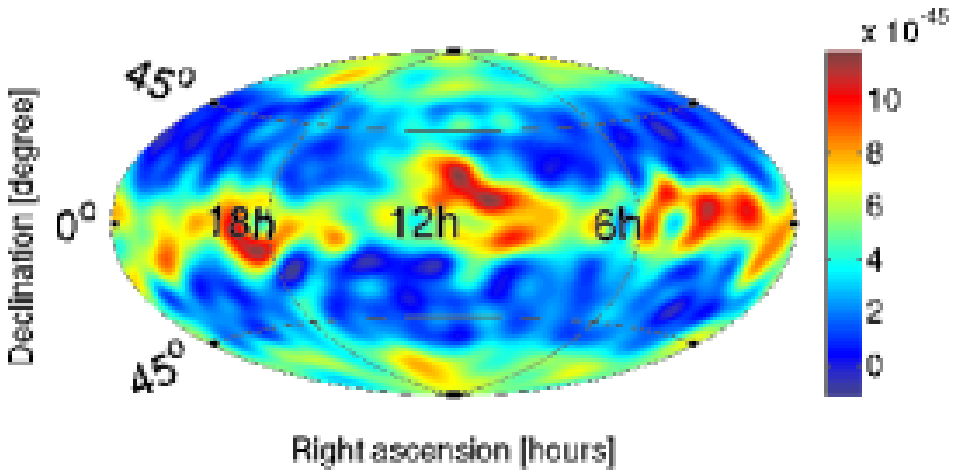,width=3in} & \psfig{file=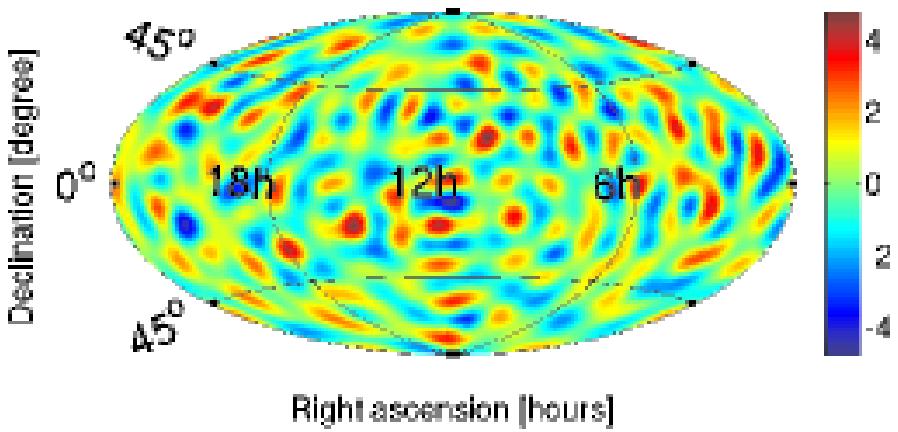,width=3in} \\
    \psfig{file=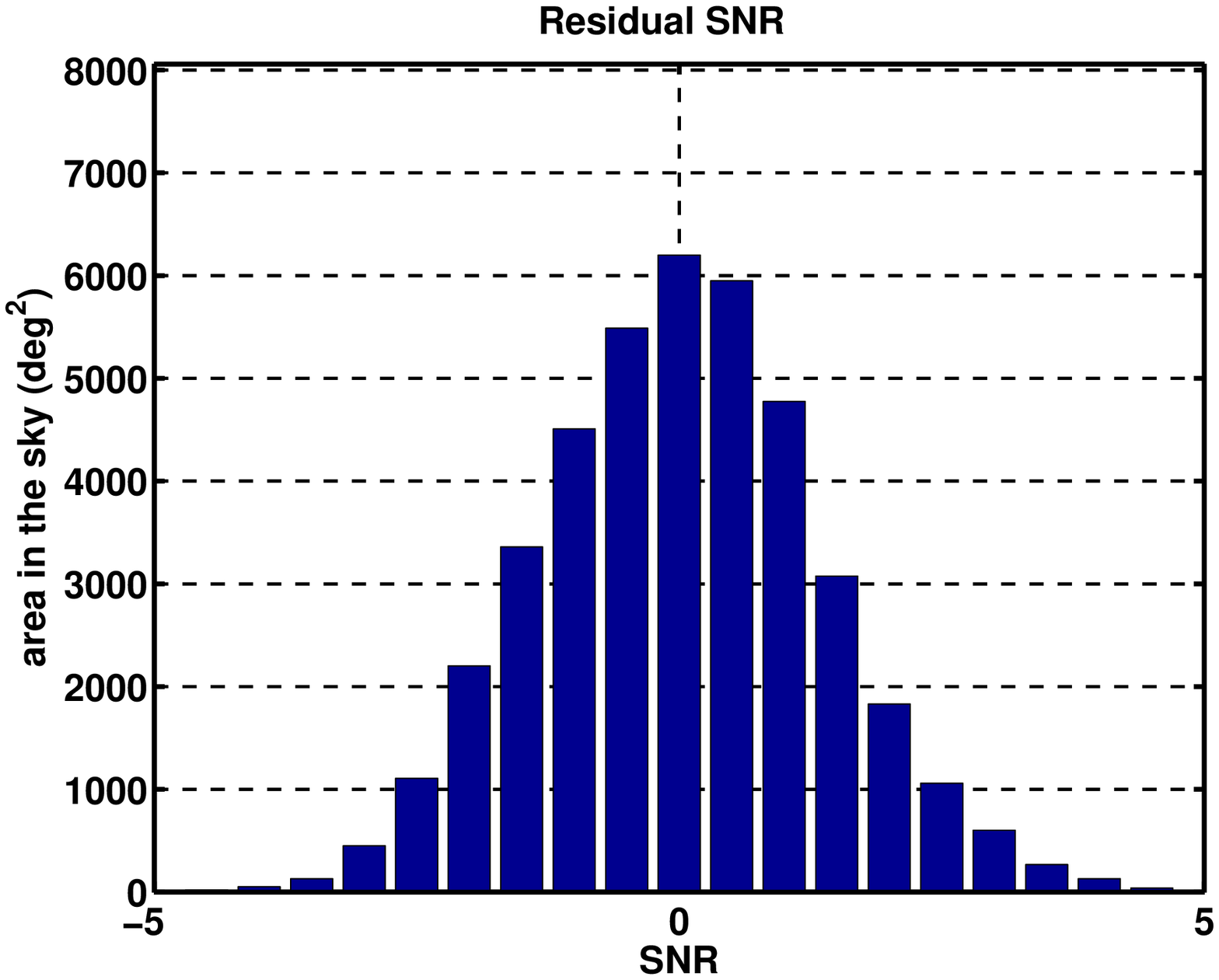,width=3in} &\\ 
  \end{tabular}
  \caption{Top-left: a toy model injection corresponding to a map measured by the WMAP satellite meant to mimic a
diffuse source clustered around ($\text{dec}=0^\circ$). The map
utilizes HEALPix~\cite{HEALPix} and the injection was simulated using
the Planck Simulator~\cite{Planck}. 
Mid-left: a regularized version of the injection.
Top right: a clean map recovered from this injection. 
Mid-right: the residuals ${\cal P}^\prime_\alpha-\hat{\cal
P}^\prime_\alpha$ normalized by $\sigma_{{\cal P}(\hat{\Omega})}$. 
Bottom-left: a histogram of these residuals.  
Note the apparent quadrupole moment present in the 
mid-left and top-right panels.
This demonstrates the relatively low sensitity to $l=2$ moments
using the H1-L1 baseline. $l_\text{max}=20$. \label{fig:gal}}
\end{figure*}

7) In the first panel of Figure~\ref{fig:gal} we plot an injection of
a diffuse source clustered about $\text{dec}=0^\circ$ generated using Planck simulator~\cite{Planck} and HEALPix~\cite{HEALPix}.
In the mid-left
panel we plot the regularized version of this injection, and 
in the top-right panel, a clean sky map using
$l_\text{max}=20$.  In the mid-right panel we 
plot ${\cal P}^\prime_\alpha-\hat{\cal P}^\prime_\alpha$.
The apparent quadrupole moment visible in
the mid-left and top-right panels illustrates the relatively low sensitity to 
$l=2$ moments 
using the H1-L1 baseline. 

\subsection{Sky maps - Multiple baselines}
\label{s:results-single-baseline}

\begin{figure}[hbtp!]
  \begin{tabular}{c}
    \psfig{file=equ_H1L1_clean_20.eps,width=3in} \\
    \psfig{file=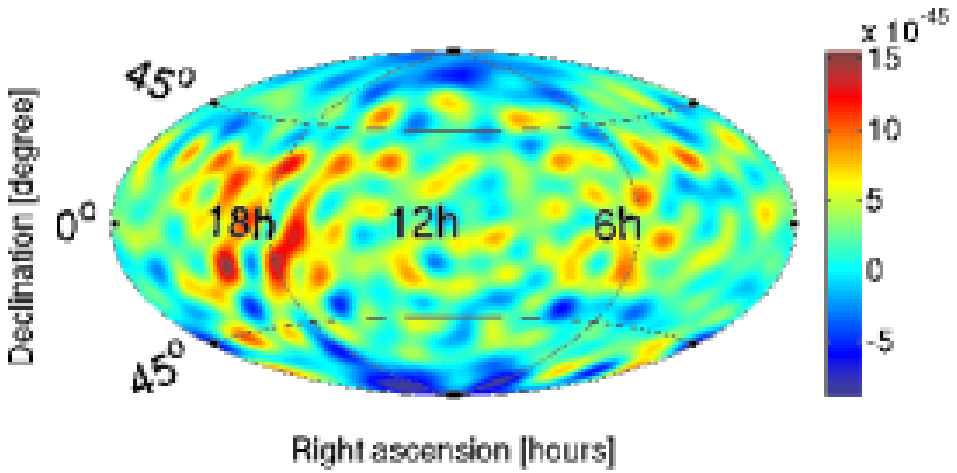,width=3in} \\
    \psfig{file=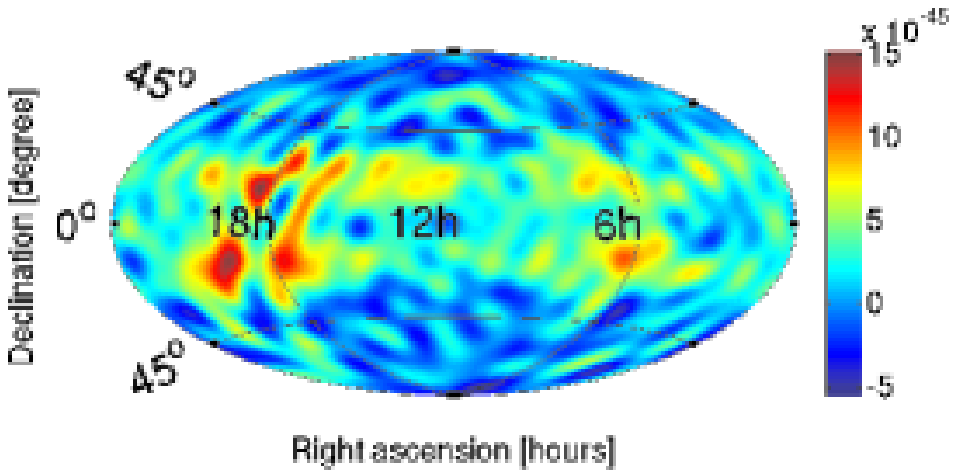,width=3in} \\
    \psfig{file=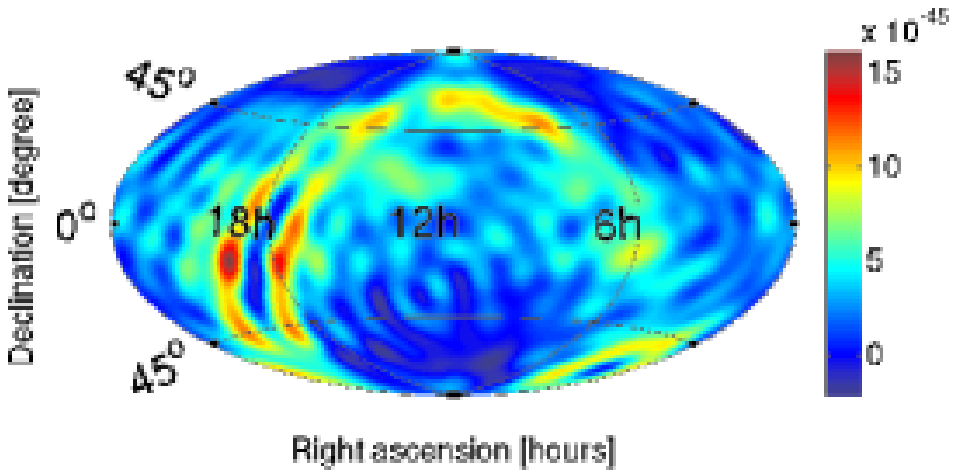,width=3in} \\ 
    \psfig{file=equatorial_injection.eps,width=3in}
  \end{tabular}
  \caption{Results from a multiple baseline simulation corresponding
  to the injection in the top panel of Figure~\ref{fig:eq}. In the top
  panel is a clean map from the H1-L1 baseline.  Second from the top
  is a clean map from the H1-V1 baseline.  
  Third is a clean map from the L1-V1 baseline.  
  Fourth is a clean map produced from combining all
  three baselines (H1-L1-V1.)  The final panel is the injected source.
  For all maps $l_\text{max}=20$.  \label{fig:multi}}
\end{figure}

Figure~\ref{fig:multi} shows the clean sky maps for a diffuse source
distributed along the galactic plane ($b=0^\circ$) and 
Figure~\ref{fig:gal} shows the clean sky maps for a diffuse source distributed along $\text{dec}=0$ analyzed with single baselines (H1-L1), (H1-V1),
and (L1-V1), and with the multi-baseline analysis (H1-L1-V1).

\section{SUMMARY}
\label{s:summary}

We have presented here a maximum-likelihood analysis
method for estimating the angular distribution of power 
in an anisotropic stochastic gravitational-wave background.
The basic idea was to cross-correlate data from a network 
of two or more gravitational-wave detectors, exploiting
time-of-arrival differences and the diurnal modulation
due to the Earth's rotation.
We derived maximum-likelihood estimators for the 
angular distribution of gravitational-wave power 
${\cal P}(\hat\Omega)=\sum_\alpha {\cal P}_\alpha
\mathbf{e}_\alpha(\hat\Omega)$, decomposed with respect
to {\em any} set of basis functions on the sky.
We derived an expression for the beam 
pattern matrix $\Gamma_{\alpha\beta}$ and discussed its
relationship to the covariance matrix 
of the maximum-likelihood estimators $\hat{\cal P}_\alpha$.
We described how singular value decomposition can be
used to regularize the inverse of $\Gamma_{\alpha\beta}$,
which was needed to remove the smearing effects of the beam 
pattern matrix on the measured (`dirty') sky maps $X_\alpha$.
We also explained how the 
single-baseline (two-detector) cross-correlation analysis 
can be extended to a network of three or more detectors, 
thereby increasing our sensitivity to detecting a signal.  
In this paper, we focused attention on a decomposition 
with respect to a basis of spherical harmonics 
$Y_{lm}(\hat\Omega)$, for which the maximum-likelihood
estimators $\hat{\cal P}_{lm}$ represent the multipole 
moments of the 
gravitational-wave sky, and for which the standard isotropic
and radiometer 
searches are recovered as special limiting cases.
Finally, we illustrated all these general results by analysing
simulated data containing injected 
stochastic gravitational-wave backgrounds having different 
angular power distributions.
 
\begin{acknowledgments}
This work was supported by NSF grants: NSF-PHY0555842, NSF-PHY-0758172, NSF-PHY-0758036, and NSF-PHY-0757058.
SM would like to acknowledge the Centre National d'\'Etudes Spatiales (France) for supporting part of the research. 
Part of the research described in this paper was carried out at the Jet Propulsion Laboratory, California Institute of Technology, under a contract with the National Aeronautics and Space Administration.

This paper has been assigned LIGO document number LIGO-P0900083.
\end{acknowledgments}

\begin{appendix}

\section{SPHERICAL HARMONICS}
\label{s:Ylms}

Our convention for the spherical harmonics $Y_{lm}(\theta,\phi)$
follow~\cite{jackson}.
Explicitly,
\begin{equation}
Y_{lm}(\theta,\phi) = \sqrt{\frac{2l+1}{4 \pi} \frac{(l-m)!}{(l+m)!}}
P_l^m(\cos \theta) e^{i m \phi}
\,,
\end{equation}
where $P_l^m(\cos\theta)$ are the associated
Legendre functions defined by
\begin{eqnarray}
&&P_l^m(x) = \frac{(-1)^m}{2^l l!} (1-x^2)^{m/2}
\frac{d^{l+m}}{dx^{l+m}} (x^2-1)^l
\,,
\\
&&P_l^{-m}(x) = (-1)^m \frac{(l-m)!}{(l+m)!} P_l^m(x)
\,.
\end{eqnarray}
The normalisation constants have been chosen so that
\begin{equation}
\int_{-1}^{1} dx\,
P_{l'}^m(x) P_l^m(x) =
\frac{2}{2 l +1}  \frac{(l+m)!}{(l-m)!} \delta_{l'l}
\end{equation}
and
\begin{equation}
\int_0^{2 \pi} d\phi \int_0^{\pi} \sin \theta d\theta \,\,
Y_{l'm'}^*(\theta,\phi) Y_{lm}(\theta,\phi) = 
\delta_{l'l} \delta_{m'm}
\,.
\end{equation}
Note that
\begin{equation}
\label{eqYcc}
Y_{l, -m}(\theta,\phi) = (-1)^m Y_{lm}^*(\theta,\phi)
\end{equation}
and
\begin{eqnarray}
\label{eqYmo}
Y_{lm}(- \hat\Omega) 
&=& Y_{lm}(\pi - \theta,\phi + \pi) 
\\
&=& (-1)^l Y_{lm}(\theta,\phi)
\\
&=& (-1)^l Y_{lm}(\hat\Omega)
\,.
\end{eqnarray}

Expressions for the first few spherical harmonics 
(up to $l=2$) are given below:
\begin{equation}
Y_{00}(\theta,\phi) = \sqrt{\frac{1}{4 \pi}}
\end{equation}

\begin{equation}
Y_{11}(\theta,\phi) = - \sqrt{\frac{3}{8 \pi}} \sin \theta e^{i \phi}
\end{equation}
\begin{equation}
Y_{10}(\theta,\phi) = \sqrt{\frac{3}{4 \pi}} \cos \theta 
\end{equation}
\begin{equation}
Y_{1,-1}(\theta,\phi) = \sqrt{\frac{3}{8 \pi}} \sin \theta e^{-i \phi}
\end{equation}

\begin{equation}
Y_{22}(\theta,\phi) = \frac{1}{4} \sqrt{\frac{15}{2 \pi}} \sin^2 
\theta e^{2 i \phi}
\end{equation}
\begin{equation}
Y_{21}(\theta,\phi) = - \sqrt{\frac{15}{8 \pi}} \sin \theta \cos \theta 
e^{i \phi}
\end{equation}
\begin{equation}
Y_{20}(\theta,\phi) = \sqrt{\frac{5}{4 \pi}} 
\left(
\frac{3}{2} \cos^2 \theta -\frac{1}{2}
\right)
\end{equation}
\begin{equation}
Y_{2,-1}(\theta,\phi) = \sqrt{\frac{15}{8 \pi}} 
\sin \theta \cos \theta e^{-i \phi}
\end{equation}
\begin{equation}
Y_{2,-2}(\theta,\phi) = \frac{1}{4} \sqrt{\frac{15}{2 \pi}} 
\sin^2 \theta e^{-2 i \phi}
\,.
\end{equation}


\section{USEFUL IDENTITIES}
\label{s:identities}

The transformation property of the spherical harmonics 
(\ref{eqYcc}) and (\ref{eqYmo})  
imply the following transformation property for the
$\gamma_{lm}$:
\begin{equation}
\label{eq:g1}
\gamma^*_{lm}(f,t) 
= 
(-1)^{l+m} \gamma_{l,-m}(f,t)
\end{equation}
and
\begin{eqnarray}
\gamma_{lm}(-f,t) 
&=& (-1)^{l} \gamma_{l,m}(f,t) 
\label{eq:g2}
\\
&=& (-1)^{m} \gamma^*_{l,-m}(f,t)\,.
\label{eq:g3}
\end{eqnarray}
Similarly, the requirement that ${\cal P}(\hat\Omega)$ is real 
implies
\begin{equation}
{\cal P}^*_{lm} = (-1)^{m} {\cal P}_{l,-m}\,.
\end{equation}
Finally, Eqs.~(\ref{eq:g1}), (\ref{eq:g2}) and (\ref{eq:g3}), 
together with the definition (\ref{e:X}) and (\ref{e:Gamma})
imply
\begin{equation}
\label{eq:Xid1}
X^*_{lm}=(-1)^m X_{l,-m} 
\end{equation}
and
\begin{equation}
\label{eq:Xid2}
\Gamma_{lm,l'm'} = 0 \;\;\;\;\;\;\;\;\; {\rm for\;odd\;} (l+l') 
\end{equation}
and
\begin{equation}
\label{eq:Xid3}
(-1)^{m+m'} \Gamma_{l,-m,l',-m'} = \Gamma^*_{lm,l'm'}  = \Gamma_{l'm',lm}
\,,
\end{equation}
so $\Gamma_{lm,l'm'}$ is Hermitian.

\section{Detection statistic}
\label{s:DetStat}


In addition to estimating the individual components 
of an anisotropic stochastic background,
it is also possible to construct a statistic that is 
optimal for detecting the presence of a background
characterized by a {\em particular} set of 
(normalized) angular components $\bar{\cal P}_\alpha$
and spectral shape $\bar H(f)$.
To show this, we note that in the presence of a signal, 
the components $X_\alpha$ of 
the dirty map can be written in the form \cite{mitra-et-al} 
\begin{equation}
X_\alpha = \Gamma_{\alpha\beta} {\cal P}_\beta + N_\alpha
\,,
\end{equation}
where $\Gamma_{\alpha\beta}$ and ${\cal P}_{\beta}$ 
are as before, and $N_\alpha$ is an additive noise term
composed of cross-correlated detector noise and stochastic
signal components.
In the weak-signal approximation, the variance of the noise-noise
cross-term, $\tilde n^*_1(f,t)\tilde n_2(f,t)$, is much greater than
that of the signal-noise cross terms, 
$\tilde h^*_1(f,t)\tilde n_2(f,t)$ and
$\tilde h^*_2(f,t)\tilde n_1(f,t)$, so to a good
approximation
\begin{equation}
N_{\alpha}
\approx
\sum_t 
\sum_f
\gamma_{\alpha}^*(f,t)
\frac{\bar H(f)}{P_1(f,t) P_2(f,t)}\,
\frac{2}{\tau}\,
\tilde n^*_1(f,t)\tilde n_2(f,t)\,.
\end{equation}
Furthermore,
when the detector noise is Gaussian and uncorrelated---an 
assumption that is well-approximated in practice---the 
$N_\alpha$ are Gaussian-distributed 
with covariance matrix
\begin{equation}
\label{e:dirtyMapNoise}
\langle N_\alpha N_\beta^* \rangle 
-\langle N_\alpha\rangle
 \langle N_\beta^*\rangle
\approx
\Gamma_{\alpha\beta}
\,.
\end{equation}

To construct the detection statistic, we assume
that the stochastic background has spectral
shape $\bar H(f)$ and {\em normalized} angular 
components $\bar{\cal P}_\alpha$ satisfying
\begin{equation}
\Gamma_{\alpha\beta}\bar{\cal P}^*_{\alpha}\bar{\cal P}_{\beta}=1
\,. 
\end{equation}
The overall amplitude $\epsilon$ of the 
background is given by
$\mathcal{P}_\alpha=\epsilon\bar{\mathcal{P}}_\alpha$.
Then the probability density function for the 
${X_\alpha}$ in the presence of such a background 
is given by the likelihood
\begin{eqnarray}
  p(\{X_\alpha\}|\epsilon)
  \propto
  \exp\bigg[-\frac{1}{2}
    (X_\alpha-\epsilon\Gamma_{\alpha\gamma}\bar{\cal P}_\gamma)^*
    (\Gamma^{-1})_{\alpha\beta}
    \nonumber
    \\
    (X_\beta-\epsilon\Gamma_{\beta\delta}\bar{\cal P}_\delta)
    \bigg]
  \,.
\end{eqnarray}
%
%
%
By the Neyman-Pearson criterion, the optimal detection 
statistic $\lambda$ is simply 
the maximum-likelihood estimator of $\epsilon$ \cite{Helstrom}---that 
is,
\begin{equation}
\lambda\equiv\hat\epsilon
\,,
\quad
\frac{d}{d\epsilon}
p(\{X_\alpha\}|\epsilon)\bigg|_{\epsilon=\hat\epsilon}
=0
\,.
\end{equation}
The result, after a straightforward calculation is
\begin{equation}
\label{eq:lambdastat}
\lambda
=
X_{\alpha}\bar{\cal P}^{*}_{\alpha}
\,,
\end{equation}
which has the form of a standard matched-filter.
Note that the detection statistic $\lambda$ 
has zero mean and unit variance in the absence 
of a signal. 
In the presence of a signal whose parameters exactly
match those of the signal model $\bar{\cal P}_\alpha$
and $\bar H(f)$, the expectation
value of the statistic is
\begin{equation}
\langle\lambda\rangle = \epsilon
\,.
\label{e:lambda}
\end{equation}
(The variance of the statistic is still unity
in the weak-signal approximation.)
In the special case of an isotropic background,
$\lambda= \hat{\cal P}_{00}/\sigma_{00}$, 
which is the signal-to-noise ratio for the 
standard isotropic search.

\end{appendix}

\bibliography{paper}

\end{document}